\documentclass[iop, revtex4, apj, numberedappendix]{emulateapj}

\newcommand{\beq}    {\begin{equation}}
\newcommand{\eeq}    {\end{equation}}
\newcommand{\beqa}   {\begin{eqnarray}}
\newcommand{\eeqa}   {\end{eqnarray}}

\newcommand{\lsim}   {\mbox{$_<\atop^{\sim}$}}

\makeatletter
\def\@seccntformat#1{\@ifundefined{#1@cntformat}%
   {\csname the#1\endcsname\quad}  
   {\csname #1@cntformat\endcsname}
}
\let\oldappendix\appendix 
\renewcommand\appendix{%
    \oldappendix
    \newcommand{\section@cntformat}{\appendixname~\thesection\quad}
}
\makeatother

\begin{document}

\title{Determination of the Cosmic Infrared Background from {\it COBE}/FIRAS and {\it Planck} HFI Observations}

\author{
N. Odegard\altaffilmark{1},
J. L. Weiland\altaffilmark{2},
D. J. Fixsen\altaffilmark{3,4},
D. T. Chuss\altaffilmark{5},
E. Dwek\altaffilmark{4},
A. Kogut\altaffilmark{4},
E. R. Switzer\altaffilmark{4}
}
\altaffiltext{1}{ADNET Systems, Inc., Code 665, NASA Goddard Space Flight Center, Greenbelt, MD 20771, USA; Nils.Odegard@nasa.gov}
\altaffiltext{2}{Department of Physics and Astronomy, Johns Hopkins University, 3400 N. Charles St., Baltimore, MD, 21218, USA}
\altaffiltext{3}{University of Maryland, College Park, MD 20742}
\altaffiltext{4}{Code 665, NASA Goddard Space Flight Center, Greenbelt, MD 20771, USA.}
\altaffiltext{5}{Department of Physics, Villanova University, 800 E. Lancaster Ave., Villanova, PA, 19085, USA.}

\begin{abstract}
New determinations are presented of the cosmic infrared background monopole brightness in the
{\it Planck} HFI bands from 100 to 857 GHz. {\it Planck} was not
designed to measure the monopole component of sky brightness, so cross-correlation of the 2015 HFI
maps with {\it COBE}/FIRAS data is used to recalibrate the zero level of the HFI maps. For the HFI 
545 and 857 GHz maps, the brightness scale is also recalibrated. Correlation of the
recalibrated HFI maps with a linear combination of Galactic H~I and H$\alpha$ data is used to 
separate the Galactic foreground emission and determine the cosmic infrared background brightness
in each of the HFI bands. We obtain CIB values of $0.007 \pm 0.014$, $0.010 \pm 0.019$, $0.060 \pm 0.023$,
$0.149 \pm 0.017$, $0.371 \pm 0.018$, and $0.576 \pm 0.034$ MJy sr$^{-1}$ at 100, 143, 217, 353, 545, 
and 857 GHz, respectively. The estimated uncertainties for the 353 to 857 GHz bands are about 3 to 6 times 
smaller than those of previous direct CIB determinations at these frequencies. Our results are compared
with integrated source brightness results from selected recent submillimeter and millimeter wavelength 
imaging surveys.
\end{abstract}

\keywords{cosmic background radiation -- cosmology: observations -- diffuse radiation -- ISM: general -- submillimeter: diffuse background -- submillimeter: ISM}

\section{Introduction}

The extragalactic background light (EBL) provides a measure of the cumulative energy
release over the history of the universe. It includes the integrated radiation
from extragalactic sources and any diffuse emission except for the cosmic
microwave background (CMB).  A significant fraction of the EBL is observed at
infrared wavelengths and is referred to as the cosmic infrared background.
At far-infrared and longer wavelengths it largely consists of radiation that
originated from stars and active galactic nuclei and has been absorbed and reradiated
by dust. Measurements of the spectrum of the cosmic infrared background monopole 
(hereafter referred to as the CIB) give important information on the cosmic
history of formation and evolution of galaxies and of their stars, black holes, and
interstellar dust.  Comparison of CIB measurements with the integrated brightness
of resolved sources from deep imaging surveys gives limits on the contribution of any
diffuse emission to the CIB and can show whether all significant contributing source 
populations have been identified.

Direct determinations of the CIB at far-infrared to millimeter wavelengths have been
made using absolute photometry from cryogenically cooled instruments with calibrated 
zero levels (see the reviews of Hauser and Dwek 2001, Kashlinsky 2005, Dwek and 
Krennrich 2013, Cooray 2016).  
Results have been obtained using {\it COBE} Far Infrared Absolute Spectrophotometer
 (FIRAS) observations from 125 $\mu$m to 2 mm 
(Puget et al. 1996, Fixsen et al. 1998, Lagache et al. 1999, Lagache et al. 2000), 
{\it COBE}/DIRBE observations at 60, 100, 140 and 240 $\mu$m (Hauser et al. 1998,
Schlegel, Finkbeiner, and Davis 1998, Lagache et al. 1999, Finkbeiner, Davis, and Schlegel 2000, Lagache et al.
2000, Wright 2004, Dole et al. 2006, Odegard et al. 2007), ISOPHOT observations from
150 to 180 $\mu$m (Juvela et al. 2009), AKARI observations at 65, 90, 140, and 
160 $\mu$m (Matsuura et al. 2011), and Spitzer MIPS total power mode observations at 
160 $\mu$m (P{\'e}nin et al. 2012). These showed that the CIB brightness peaks at about
200 $\mu$m and that the integrated CIB brightness from 10 to 1000 $\mu$m is comparable to 
that of the cosmic optical background from 0.1 to 10 $\mu$m. 

Deep source surveys have resolved 
a large fraction of the CIB at far-infrared to millimeter wavelengths into 
discrete sources that are predominantly dusty star forming galaxies at $z \gtrsim 1$ (for
reviews see Lagache, Puget, and Dole 2005 and Casey, Narayanan, and Cooray 2014).
As wavelength increases, the relative contributions of higher-redshift sources
and cooler sources to the CIB increase (e.g., Zavala et al. 2017).

The accuracy of the direct CIB determinations is limited by the accuracy at which Galactic
and interplanetary foregrounds can be modeled and subtracted from the data, and in some
cases by the accuracy of zero-level calibration. Quoted uncertainties are typically 20\% to 30\%.
For example, the amplitude uncertainty of the widely adopted 125 $\mu$m to 2 mm 
CIB spectrum of Fixsen et al. (1998) is $\pm$ 30\%.  This uncertainty
limits our knowledge of the fraction of the CIB that has been resolved by source surveys.

Other determinations of the CIB have made use of CIB anisotropy measurements.
P{\'e}nin et al. (2012) scaled their direct 160 $\mu$m CIB determination to 100 $\mu$m assuming
that the 100 $\mu$m/160 $\mu$m color they measured for CIB fluctuations is valid for the 
CIB monopole. 
Planck Collaboration 2013 Results XXX (2014) obtained CIB estimates in five bands from
100 $\mu$m to 1380 $\mu$m using an extended halo model to extrapolate CIB power spectra
determined from {\it IRAS} and {\it Planck} data to multipole $l$=0.
Both of these analyses assume that any contribution of diffuse emission to the CIB
is negligible.

In this paper, we make new, more accurate determinations of the CIB at these wavelengths using
{\it COBE}/FIRAS and {\it Planck} High Frequency Instrument (HFI) data. In \S2 we describe our 
use of FIRAS data to recalibrate the zero levels and gains of 2015 data release HFI maps.
Most of the work reported here was done before the 2018 {\it Planck} data were released. Differences 
between the 2018 and 2015 intensity maps are small, typically at most 2-3\% (Planck Collaboration 
2018 Results IV 2018), so we do not expect that use of the 2018 maps would give substantially different
CIB results. An attempt by the {\it Planck} team to calibrate 2013 HFI data using FIRAS data 
(Planck Collaboration 2013 Results VIII 2014) gave problematic results that the {\it Planck} team did not
adopt or investigate further in later work and that we do not find from our analysis. In \S3 we present
correlations of the recalibrated HFI maps with Galactic H~I 21-cm line emission, and with a linear
combination of Galactic H~I and Galactic H$\alpha$ emission. These are used to 
separate the Galactic foreground emission and determine a CIB value in each of the HFI bands.
The high angular resolution and sensitivity of the HFI data allow the correlations to be 
established more accurately and to lower H~I column density than was possible with the 7$\arcdeg$ 
resolution FIRAS data, resulting in significant improvement in the accuracy of the derived CIB. 
We also present an alternative method of CIB determination for the HFI 100 to 545 GHz bands in which the
recalibrated CIB-subtracted 857 GHz map is used as a Galactic foreground emission template. 
In \S4 we compare our CIB results with the integrated brightness of sources from ALMA, BLAST, 
{\it Herschel}/SPIRE, SCUBA, and SCUBA-2 observations. Our results are summarized in \S5.

\section{Recalibration of HFI Maps}

{\it Planck} was not designed to measure the monopole component of sky brightness.
The zero points of the HFI maps in the {\it Planck} 2015 data release were set by a 
two-step process (Planck Collaboration 2015 Results VIII 2016): (1) The HFI map from 
each detector was correlated with observations of Galactic H~I emission for sky
regions of low H~I column density where H$_2$ is negligible.  A linear fit to the
correlation was extrapolated to zero H~I column density to set the map zero level
for Galactic emission. (2) A CIB monopole brightness calculated from the empirical
galaxy evolution model of B{\'e}thermin et al. (2012b) was then added to the map. The
uncertainty of the model CIB prediction was estimated to be 20\%.
The gain calibration was based on observations of the
orbital CMB dipole in the 100 to 353 GHz bands and observations of Uranus and Neptune
in the 545 and 857 GHz bands.

We use FIRAS destriped sky spectra to recalibrate the zero levels of the HFI maps in
all bands and to recalibrate the brightness scale of the HFI 545 GHz and 857 GHz maps.
Accurate calibration of the zero level and gain for the FIRAS data was established
by interspersing sky measurements with measurements of an external blackbody calibrator
that filled the FIRAS beam. We describe our preparation of the HFI and FIRAS data to produce
maps that have common beam response, frequency response, and pixelization, with common 
CMB dipole and zodiacal light subtraction. We also describe our subtraction of the CMB 
monopole from the FIRAS data. We then present results of linear fits to the 
FIRAS-HFI correlation for each HFI band, which we use to recalibrate the HFI maps.

\subsection{{\it Planck} HFI Data}

We use {\it Planck} full-mission, full-channel Stokes I maps for the HFI frequency bands
from data release 2.02, which were made public in 2015 at the Planck Legacy Archive. 
These maps have had the {\it Planck} 2015 'nominal' CMB dipole subtracted, and have had
zodiacal light subtracted using a fit of the Kelsall et al. (1998) interplanetary dust cloud
model to 2015 HFI data 
(Planck Collaboration 2013 Results XIV 2014, Planck Collaboration 2015 Results VIII 2016).
The released 545 GHz and 857 GHz band maps are maps of intensity in MJy sr$^{-1}$ at 
the nominal band frequency under the assumption that the spectrum $\nu I_\nu$ is constant across
the bandpass. For the 100 GHz to 353 GHz bands, we use conversion factors provided by the {\it Planck} 
team to convert the released maps from thermodynamic temperature in K to MJy sr$^{-1}$ for
constant $\nu I_\nu$. Thus, the intensity in each HFI band map we work with is
\begin{equation}
I^{HFI}_{\nu_0} = \frac {\int I_\nu R_\nu \thinspace d\nu} {\int (\nu_0/\nu)  R_\nu \thinspace d\nu}, 
\end{equation}
where $\nu_0$ is the nominal frequency and $R_\nu$ is the HFI band frequency response function.

We smoothed each HFI map with the instantaneous FIRAS beam (Brodd et al. 1997) and then smoothed
in ecliptic latitude with a 2.4 degree boxcar to account for FIRAS scan motion
during the integration of a FIRAS interferogram (Fixsen et al. 1997). Each map was then 
degraded to HEALPix $N_\mathrm{side} = 256$\footnote{A HEALPix map is divided into 12$N_\mathrm{side}^2$
pixels, with each pixel of width $\sim58.6\arcdeg/N_\mathrm{side}$. See G{\'o}rski et al. (2005)
and \url{http://healpix.sourceforge.net.}} (pixel size $0.23 \arcdeg$) using flat weighting
and averaged over FIRAS pixels 
(pixel size $2.6\arcdeg$) in the {\it COBE} quadrilateralized spherical cube projection 
(Brodd et al. 1997).
The available FIRAS interferograms do not give dense uniform sampling for all pixels, so
the mean observed position for a pixel can be significantly offset from the pixel center.
As a final step, we interpolated between pixels in each map to obtain the smoothed HFI brightness
value at the mean position observed by FIRAS for each pixel.

The {\it Planck} HFI observations were divided into five surveys, each covering about 6 months.
We estimated uncertainties for the smoothed full-mission HFI data from the variation among
smoothed individual survey maps. 
For each band, we formed the difference of each survey map from the full-mission
map and smoothed and pixelized it as described above. We then calculated the $1\sigma$ uncertainty
of the smoothed full-mission map for each FIRAS pixel as the standard error of
the mean for the difference maps from the surveys in which that pixel was fully sampled.
These uncertainty estimates include contributions from destriper uncertainties and far
sidelobe uncertainties. As discussed below, they are negligible in comparison to the 
uncertainties for the FIRAS data. 
  
\subsection{{\it COBE} FIRAS Data}

We use data from the final (pass 4) {\it COBE} project release of the FIRAS low spectral 
resolution destriped sky spectra, which cover 99\% of the sky over frequency channels from 
68 to 2911 GHz with a channel spacing of 13.6 GHz. Descriptions of the FIRAS instrument,
data calibration, data processing, and pass 4 data products are given in the FIRAS 
Explanatory Supplement (Brodd et al. 1997). We combine data from the 
FIRAS low-frequency (LOWF) and high-frequency (HIGH) destriped sky spectra datasets
for the FIRAS pixels that contain data in each dataset. The two datasets share 
three channels in common.  For each of these channels we use the dataset with the 
lower detector noise, the LOWF data for the 612.2 and 625.8 GHz channels and the 
HIGH data for the 639.4 GHz channel.

To prepare for correlation with the smoothed HFI map for each HFI band, we subtract CMB monopole,
CMB dipole, and zodiacal light contributions and form a weighted average over frequency
channels using weights to match the HFI band frequency response as closely as
possible. Color correction is then applied to account for the difference in frequency response
between the HFI band data and the channel-averaged FIRAS data. A more detailed
description of this processing follows.

\subsubsection{CMB Monopole Subtraction}

We subtract the isotropic CMB blackbody spectrum from the FIRAS data for each pixel
using a best-fit CMB temperature for the pass 4 FIRAS data of 2.72765 K. 
We use this best-fit temperature rather than the latest CMB temperature determination of
2.72548 $\pm$ 0.00057 K (Fixsen 2009), because there is a known bias in the pass 4 data due to 
an error in the original temperature determination for the FIRAS external calibrator 
(Mather et al. 1999, Fixsen 2009).

\begin{figure*}
\figurenum{1}
\epsscale{1.00}
\plotone{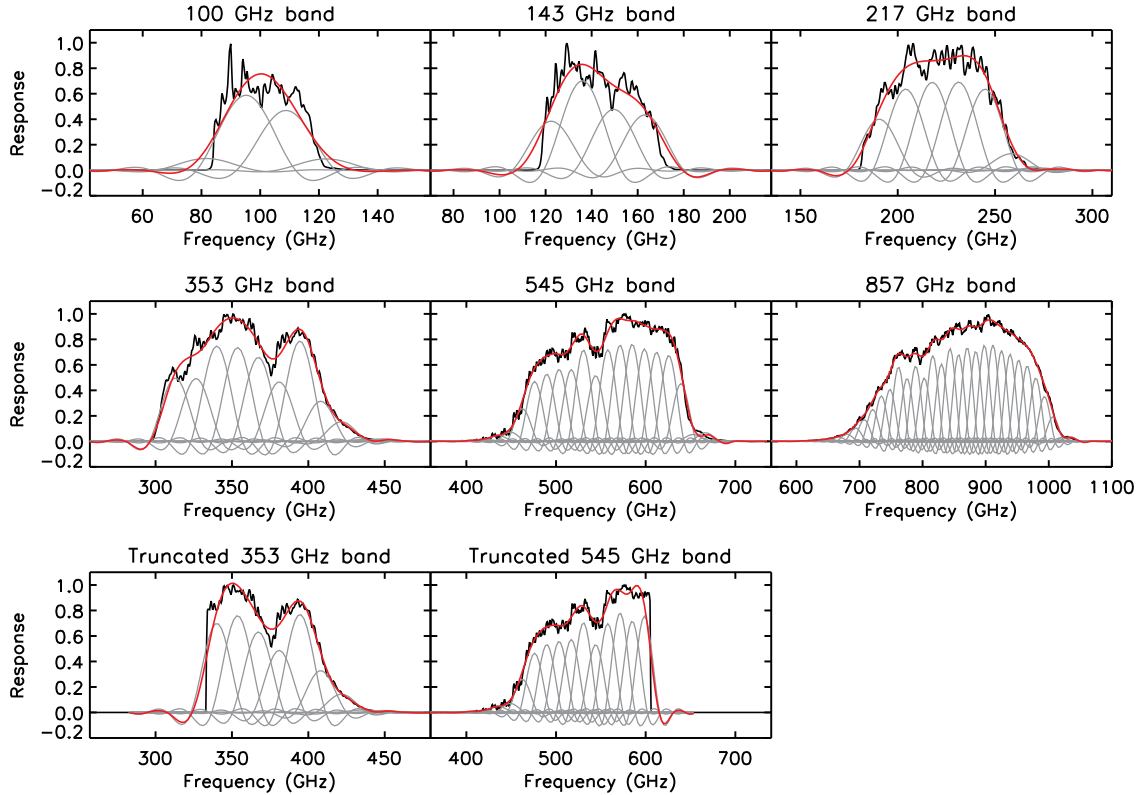}
\caption{{\it Planck} HFI bandpasses (black) and effective bandpasses for 
channel-averaged FIRAS data (red) formed by fitting a linear combination
of frequency response functions for individual FIRAS channels (gray).
The last two panels show fits to truncated versions of the HFI 353 GHz and
545 GHz bandpasses, which provide alternate effective FIRAS bandpasses 
that exclude questionable FIRAS channels. Color correction is applied to
the channel-averaged FIRAS data to correct for the difference between the
effective FIRAS bandpass and the HFI bandpass.}
\end{figure*}

We obtained the best-fit temperature from fits of a CMB blackbody spectrum and different Galactic
dust emission models to FIRAS pass 4 destriped sky spectra from 100
to 2100 GHz and DIRBE 100 $\mu$m and 240 $\mu$m data. Before the fitting, the CMB dipole was 
subtracted from the data using the dipole amplitude and direction from Hinshaw et al. (2009), 
and estimates of the CIB monopole, zodiacal light, Galactic free-free, 
and Galactic synchrotron contributions were subtracted as described in Appendix A. The data 
were then averaged over each of six large
sky regions that exclude the Galactic plane.  The fits were done as described by Odegard 
et al. (2016) and used either the dust emission model of Meisner and Finkbeiner (2015, MF) or the 
two-level systems (TLS) dust model of Meny et al. (2007) and Paradis et al. (2011). These models 
both give acceptable quality fits, but the TLS model predicts progressive flattening of
the dust spectrum with decreasing frequency below 300 GHz that is not predicted by the MF model.

We adopt an uncertainty of 46 $\mu$K for the best-fit CMB temperature. The uncertainty calculation
is described in Appendix A. The uncertainty is dominated by the contribution from uncertainty in 
the CIB monopole spectrum subtraction, but also includes contributions from FIRAS and DIRBE 
measurement uncertainties and from differences in the fit results for the different dust emission
models. We do not include a 
contribution from the uncertainty in the absolute temperature scale of the FIRAS external calibrator 
(PTP uncertainty, Brodd et al. 1997). This is a systematic uncertainty that does not
affect the precision of subtracting the best-fit CMB monopole signal from the data.
It only affects the accuracy of absolute CMB temperature determination.

\begin{deluxetable}{cc}
\scriptsize
\tablewidth{0pt}
\tablecaption{CMB Monopole Subtraction Uncertainties\label{tab:cmb_mono_unc_tab}}
\tablehead{
\colhead{HFI Band} &
\colhead{CMB Subtraction Uncertainty}\\
\colhead{} &
\colhead{(MJy sr$^{-1}$)}
}
\startdata
100 GHz & 0.011 \cr
143 GHz & 0.017 \cr
217 GHz & 0.022 \cr
353 GHz & 0.013 \cr
545 GHz & 0.0027 \cr
857 GHz & 0.0001
\enddata
\end{deluxetable}

The CMB monopole subtraction uncertainty averaged over each HFI band is calculated from the
46 $\mu$K temperature uncertainty using the {\it Planck} team's conversion factor from thermodynamic 
temperature to MJy sr$^{-1}$. These uncertainties are given in Table 1. They are included in 
the uncertainties of the zero levels for the recalibrated HFI maps in \S2.3 and are included in
the uncertainties of our CIB determinations in \S3.4.

We note that our best-fit CMB temperature is consistent with the CMB temperature of $2.728 \pm 0.004$ K
reported by the FIRAS team from analysis of the pass 4 data (Fixsen et al. 1996). Their quoted
uncertainty (and the uncertainty of the Fixsen (2009) temperature determination given above)
allows for systematic errors including external calibrator temperature uncertainty that do not apply
for subtraction of the best-fit monopole signal from the data.

\subsubsection{Averaging over Frequency Channels}

After CMB monopole subtraction, we average over the FIRAS frequency channels in each HFI bandpass 
using weights determined by fitting a linear combination of the individual FIRAS channel
frequency response functions to the HFI band frequency response function.
The frequency response function for a FIRAS channel is calculated as the
Fourier transform of the apodization function that was applied to the co-added
FIRAS interferograms before they were Fourier-transformed into spectra. We
calculate the apodization function following section 5.1 of the FIRAS explanatory supplement 
(Brodd et al. 1997). The apodization function is padded with zeros such that the frequency
spacing of its Fourier transform is 13.6 GHz and then folded about the interferogram 
peak position before
Fourier transformation. The fits to the HFI bandpasses are shown in Figure 1. 
The channel-averaged intensity for each HFI band and each FIRAS pixel is calculated as
\begin{equation}
I^{FIRAS}_{\nu_0} = \frac {\sum\limits_{i} I^{FIRAS}(\nu_i) \thinspace A_i \thinspace \Delta \nu} {\sum\limits_{i} (\nu_0/\nu_i) \thinspace A_i \thinspace \Delta \nu}, 
\end{equation}
where $\nu_0$ is the nominal HFI band frequency, $i$ is the FIRAS channel index, $A_i$ is the channel
response function fit amplitude for channel $i$, and $\Delta \nu$ is the channel width.

There are potential FIRAS data quality issues for some of the FIRAS channels that fall within the
HFI 353 GHz and 545 GHz bandpasses.  The channels between 306 and 333 GHz may be affected by residual
mirror transport mechanism ghosts.  The channels between 605 and 687 GHz may be affected by a
significant variation of the dichroic filter frequency response across the channel, which is not 
included in our calculation of the channel frequency response, or other problems 
(Brodd et al. 1997, Finkbeiner, Davis, and Schlegel 1999).
To check for possible effects on our results, we make alternative channel averages for these bands that
exclude the questionable channels. The FIRAS data are averaged over truncated versions of the
HFI bandpasses as shown in the last two panels of Figure 1, with the 353 GHz bandpass zeroed below
333 GHz and the 545 GHz bandpass zeroed above 605 GHz.  

\subsubsection{CMB Dipole Subtraction}

We subtract the CMB dipole signal from the channel-averaged data using the {\it Planck} team's 
2015 nominal dipole amplitude and direction (3364.5 $\mu$K toward $l=264.00\arcdeg$, $b=48.24\arcdeg$,
Planck Collaboration 2015 Results I 2016),
their conversion factors from thermodynamic temperature to
MJy sr$^{-1}$ for the HFI bands,
and color corrections to scale the dipole brightness averaged over the HFI band 
frequency response to brightness averaged over the channel-averaged
FIRAS frequency response.
The frequency dependence of the CMB dipole spectrum is given by $dB_{\nu}(T_0)/dT$,
where $B_{\nu}(T_0)$ is the Planck function at the CMB monopole temperature, so
the color correction for a given band is calculated as
\begin{equation}
C_{CMB} = \frac {\int dB_{\nu}(T_0)/dT \thinspace R_{\nu}^{'} \thinspace d\nu} {\int (\nu_0/\nu) \thinspace R_{\nu}^{'} \thinspace d\nu} \frac {\int (\nu_0/\nu) \thinspace R_{\nu} \thinspace d\nu} {\int dB_{\nu}(T_0)/dT \thinspace R_{\nu} \thinspace d\nu},
\end{equation}
where $R_{\nu}^{'}$ is the frequency response function for the channel-averaged FIRAS data 
and $R_\nu$ is the frequency response function for the HFI data.

\subsubsection{Zodiacal Light Subtraction}

\begin{figure*}
\figurenum{2}
\epsscale{0.9}
\plotone{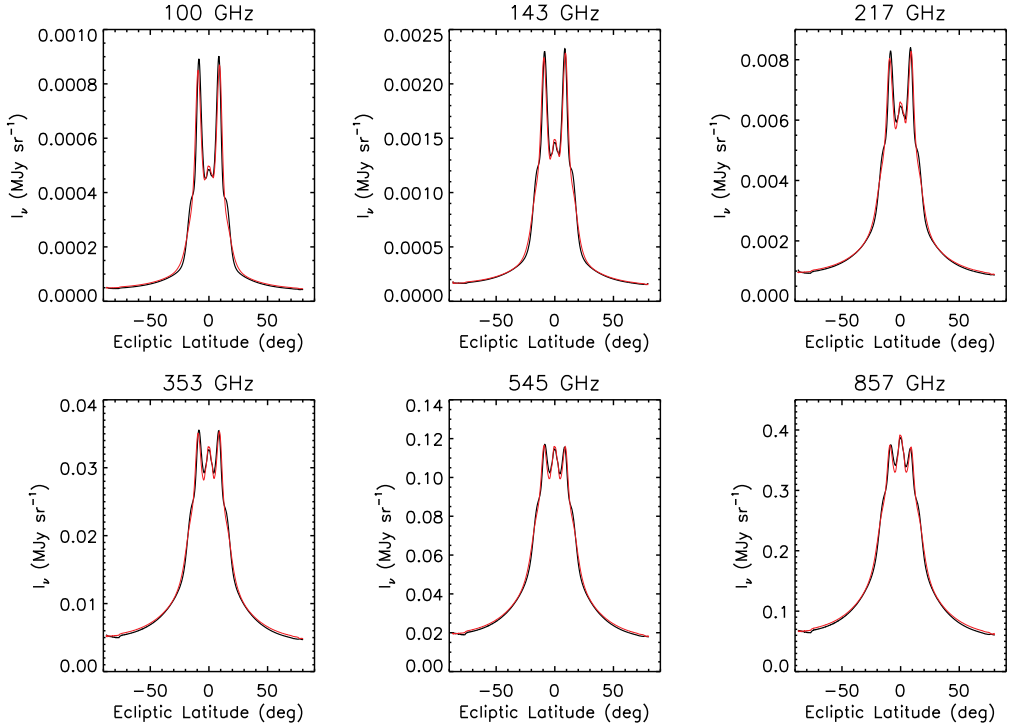}
\caption{The black curves show the ecliptic latitude profile of the {\it Planck} 2015 
survey 1 zodiacal light correction map for each HFI band, and the red curves show
results from fits of dust component template maps that we use to estimate the dust emissivities
for the {\it Planck} 2015 zodiacal light model (see text).}
\end{figure*}

The interplanetary dust emission observed for a given sky pixel depends on the spacecraft's
position
in the interplanetary dust cloud at the time of observation. To subtract it from 
the channel-averaged FIRAS data in a manner consistent with that used for the HFI 
data, we need to evaluate the model used by the {\it Planck} team at the times of FIRAS observations. 
This model was determined by fitting geometrical
dust cloud components of the Kelsall et al. (1998) interplanetary dust cloud model 
to HFI data. For each HFI band, the blackbody emissivity of the dust in each component was
adjusted to obtain the best fit. 
The dust cloud components used for the 2015 model are the diffuse dust cloud
and three sets of asteroidal dust bands (Planck Collaboration 2015 Results VIII 2016).
We have calculated predictions of this model using a modified version of the DIRBE team 
software for evaluating the Kelsall et al. model
that includes the HFI bands. We initially tried to evaluate the model for HFI observations 
using the {\it Planck} team's best-fit emissivities 
for the different cloud components given in Planck Collaboration 2015 Results VIII (2016). 
We found that this underpredicts the released {\it Planck} 2015 zodiacal light correction 
maps in the HFI bands by 20\% to 30\% at low ecliptic latitudes. 
This has been traced primarily to a problem in {\it Planck} team 2015 zodiacal model 
evaluation software, discovered after data release (K. Ganga, private communication 2015).

\begin{deluxetable}{ccccc}
\scriptsize
\tablewidth{0pt}
\tablecaption{Adopted Zodiacal Dust Emissivity Values}
\tablehead{
\colhead{HFI Band} &
\colhead{Smooth Cloud} &
\colhead{Band 1} &
\colhead{Band 2} &
\colhead{Band 3}
}
\startdata
100 GHz & 0.014 & 1.25 & 0.15 & 0.50 \cr
143 GHz & 0.023 & 1.39 & 0.22 & 0.89 \cr
217 GHz & 0.063 & 1.85 & 0.40 & 1.22 \cr
353 GHz & 0.132 & 2.41 & 0.80 & 1.96 \cr
545 GHz & 0.210 & 2.81 & 1.11 & 2.81 \cr
857 GHz & 0.285 & 3.23 & 1.58 & 3.60
\enddata
\end{deluxetable}

Instead of using the 2015 published emissivities, we estimated emissivity values for the 
{\it Planck} team model by fitting a linear combination of dust cloud component template
maps to the 2015
zodiacal light correction maps for survey 1, survey 2, and survey 3 in each HFI band.
We chose not to use the maps for survey 4 or survey 5 because the sky coverage for these surveys
is less complete.
For the template maps, we calculated maps of emission from the smooth dust cloud and
each set of dust bands for each of the three surveys, using
the published emissivity values and the released dates of observation
for each survey. The templates were calculated using the mean date of observation 
for each pixel. Pixels that were not observed in all three surveys or that were
observed over more than a 7-day period in any survey were not used in the fitting.
For a given HFI band, we made a simultaneous fit of the template maps to
the survey 1, survey 2, and survey 3 zodiacal light correction maps. For each survey,
\begin{equation}
I^{zodi}_i = \sum\limits_{j} c_j \thinspace I^{template}_{i,j},
\end{equation}
where $i$ is a survey index and $j$ is a dust cloud component index. We determined $c_j$ values that
minimize the mean square deviation of the fit from the zodiacal light correction maps.
These were used to scale the published emissivity values to obtain the values listed in Table 2. 

We used these emissivities to evaluate the zodiacal light model 
for each HFI band at the mean observation time and mean pointing direction of each 
FIRAS co-added interferogram. The results were averaged over FIRAS pixels, smoothed 
to the FIRAS beam, color-corrected
to scale from brightness averaged over the HFI band frequency response
to brightness averaged over the FIRAS frequency response, and then
subtracted from the channel-averaged data. The color correction factors were
calculated separately for each pixel using the zodiacal light spectrum derived 
from the {\it Planck} team model using our emissivity values.
The quality of the template fits is illustrated by Figure 2, which compares 
the ecliptic latitude profile of the survey 1 zodiacal light correction map 
with that of the fit for each HFI band. 

\begin{deluxetable*}{cccc}
\scriptsize
\tablewidth{0pt}
\tablecaption{Zodiacal Light Subtraction Uncertainties}
\tablehead{
\colhead{HFI Band} &
\colhead{Model Geometry Uncertainty} &
\colhead{Model Emissivity Uncertainty} &
\colhead{Total Uncertainty}\\
\colhead{} &
\colhead{(MJy sr$^{-1}$)} &
\colhead{(MJy sr$^{-1}$)} &
\colhead{(MJy sr$^{-1}$)}
}
\startdata
100 GHz & 0.00001 & 0.00002  & 0.00002 \cr
143 GHz & 0.00004 & 0.00008 & 0.00009 \cr
217 GHz & 0.0002 & 0.0008 & 0.0008 \cr
353 GHz & 0.0009 & 0.0044 & 0.0045 \cr
545 GHz & 0.0031 & 0.0089 & 0.0095 \cr
857 GHz & 0.011 & 0.001 & 0.011
\enddata
\end{deluxetable*}

Table 3 lists our estimated zodiacal light subtraction uncertainty for HFI zero-level
calibration and CIB determination for each HFI band.
We add two contributions in quadrature to obtain a total uncertainty.  We include a
contribution due to uncertainty in the geometry of the diffuse dust cloud and a contribution
due to uncertainty in the dust emissivities. The total uncertainties are included in the
uncertainties of the zero levels for the recalibrated HFI maps in \S2.3 and are included in
the uncertainties of our CIB determinations in \S3.4.

Kelsall et al. (1998) estimated uncertainties of their model predictions in the DIRBE bands
at high ecliptic latitudes by comparing results of models with different diffuse dust cloud 
geometries that gave comparable quality fits to the time variation of the DIRBE data. They
obtained an uncertainty of 0.5 nW m$^{-2}$ sr$^{-1}$ for the DIRBE 240 $\mu$m (1250 GHz) 
band.\footnote{An alternative estimate of the DIRBE 240 $\mu$m model prediction uncertainty is 
given by the difference between the DIRBE 240 $\mu$m CIB value of Hauser et al. (1998)
and that obtained by Wright (2004). The main difference between these analyses is that Hauser et al. 
subtracted zodiacal light using the Kelsall interplanetary dust model and Wright used the model of
Gorjian et al. (2000),
which included a constraint that the residual DIRBE 25 $\micron$ intensity
after zodiacal light subtraction be zero at high Galactic latitude. Hauser et al. reported
a 240 $\mu$m CIB value of 13.6 $\pm$ 2.5 nW m$^{-2}$ sr$^{-1}$ and Wright reported a value of 
13 $\pm$ 2.5 nW m$^{-2}$ sr$^{-1}$, so the difference is similar to the Kelsall et al. uncertainty 
of 0.5 nW m$^{-2}$ sr$^{-1}$.}
We extrapolated this to each of the HFI bands to obtain the model geometry uncertainties
listed in Table 3.
The extrapolation was done using the sky-averaged ratio of the {\it Planck} model prediction for the
HFI band to the Kelsall model prediction for the 240 $\mu$m band, with both predictions calculated 
for the FIRAS observing times and pointing directions. 

We note that detection of an isotropic component of zodiacal emission was
reported by Kondo et al. (2016) from their analysis of {\it AKARI} 9 $\mu$m and 18 $\mu$m maps.
Based on their estimated spectrum of this component, its brightness in the HFI bands is expected
to be more than an order of magnitude smaller than our adopted model geometry uncertainties.
Rowan-Robinson and May (2013) reported a much brighter isotropic component from their analysis
of DIRBE and {\it IRAS} data at 12, 25, 60, and 100 $\mu$m.  Extrapolating using the spectral shape 
from Kondo et al., its brightness in the HFI bands is about 0.6 of our 
adopted model geometry uncertainties.

For the contribution of dust emissivity uncertainties, we compare results obtained using our
2015 template fit emissivities in Table 2 with results obtained using the {\it Planck} 2018 emissivities,
which were released after most of the analysis reported in this paper was completed. The 2018
emissivities are expected to be more accurate due to improvements in {\it Planck} 2018 data processing.
The two sets of emissivities were used to perform two cases of zodiacal light subtraction from the
FIRAS data and to obtain two sets of HFI 2015 zero-level recalibration offsets as described in \S2.3. 
The difference between the two sets of offsets is listed as the emissivity uncertainty contribution 
in Table 3. These uncertainties are primarily due to differences between the 2015 and 2018 zodiacal 
light models, but they also account for any errors in our template fit emissivities due to contributions
other than zodiacal light to the zodiacal light correction maps. The maps include small offset 
effects due to differences in destriping between HFI maps with and without zodiacal light removal 
(Planck Collaboration 2015 Results VIII 2016). 

\subsubsection{Color Corrections}

We apply color corrections to scale the channel-averaged data after subtraction of 
CMB monopole, CMB dipole, and zodiacal
light to the HFI band frequency response. We treat this data as a sum
of CIB monopole, CMB fluctuation, and Galactic emission components and we calculate
separate color correction factors for each component. These are thought to be the
dominant remaining contributors to the diffuse emission at HFI frequencies. 
Figure 3 shows estimated contributions of the most important diffuse emission
components in the HFI bands for a sky region that excludes the Galactic plane.

\begin{figure*}
\figurenum{3}
\epsscale{0.90}
\plotone{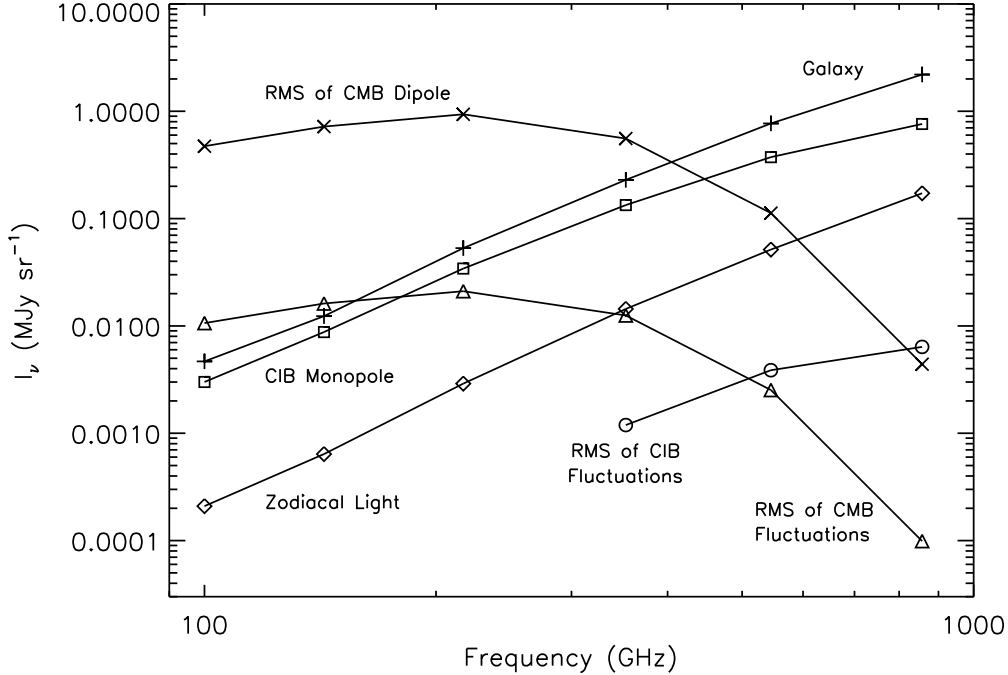}
\caption{Estimated contributions of diffuse emission components in the HFI bands. For Galactic emission
and zodiacal light, mean intensity is shown for our sky region 3 (described in \S2.3), which 
excludes the Galactic plane
and covers 64\% of the sky. For the CMB dipole, CMB fluctuations, and CIB fluctuations, RMS anisotropy
is shown for maps smoothed to FIRAS resolution ($\sim7\arcdeg$ FWHM), using the {\it Planck} 2015 
nominal CMB dipole and {\it Planck} 2015 SMICA CMB map over the full sky and the {\it Planck} 2015 
GNILC CIB anisotropy maps at $\vert b \vert > 50\arcdeg$.}
\end{figure*}

For the CIB monopole, we adopt the spectral shape $I^{CIB}_{\nu} \propto {\nu}^{1.4} B_{\nu}(13.6 K)$,
from Gispert et al. (2000) and an amplitude of 0.35 MJy sr$^{-1}$ at 545 GHz from the 
B{\'e}thermin et al. (2012b) galaxy evolution model (Planck Collaboration 2015 Results VIII 2016).
We calculate an average over FIRAS channels for each HFI band, $I^{CIB}_{\nu_0}$, as in equation (2).
For CMB fluctuations, we calculate an average over FIRAS channels $I^{CMB}_{\nu_0}$ 
for each band and each FIRAS pixel using the {\it Planck} 2015 SMICA map smoothed to the 
FIRAS beam, the conversion factor from thermodynamic temperature to MJy sr$^{-1}$ for the 
HFI band, and the HFI to FIRAS color correction $C_\mathrm{CMB}$ from equation (3). We calculate
the Galactic emission component for each band and pixel as the residual emission after 
subtraction of $I^{CIB}_{\nu_0}$ and $I^{CMB}_{\nu_0}$.

The FIRAS to HFI color corrections for the CIB component are calculated using the adopted 
CIB monopole spectrum, those for the CMB fluctuation component are 1/$C_\mathrm{CMB}$ from equation (3), 
and those for the Galactic component are
calculated using the foreground model from Planck Collaboration 2015 Results X (2016).
For the 217 GHz and higher frequency bands, the model is dominated by thermal dust emission
and we use the spectrum of this component only. For the 143 GHz band, we use the combined spectrum
of the thermal dust, free-free, and synchrotron emission components. For the 100 GHz
band, we use the combined spectrum of these components plus the 115.27 GHz CO line emission
component. The color correction values for the CIB, CMB, and Galactic components are
within 1.2\% of unity for all bands and all FIRAS pixels, due to the good agreement between the HFI and FIRAS
response functions (Figure 1). For the truncated FIRAS response functions for the 353
and 545 GHz bands, the net color correction is typically about 0.89 and 1.14, respectively.

\subsection{FIRAS-HFI correlations}

We make linear fits to correlations between the FIRAS and HFI maps processed to a common format 
as described above.
For each HFI band, we make a series of fits for five different sky regions shown in Figure 4.
Region 5 covers the entire sky except for the exclusion of pixels with FIRAS pixel weight less than 0.4 
(pixels with a small number of observations) and pixels for which the FIRAS data are contaminated by Mars 
or Jupiter.  Region 5 includes the Galactic center, so it samples the full range of sky brightness.
Region 4 excludes the same pixels as region 5 and also excludes pixels in the inner Galactic plane where 
$\vert b \vert < 10\arcdeg$ and $l < 120\arcdeg$ or $l > 260\arcdeg$. 
Regions 1, 2, and 3 exclude the same pixels as region 4 and also exclude pixels
where the smoothed HFI 857 GHz map is brighter than 2.5, 5, and 10 MJy sr$^{-1}$, respectively.

\begin{figure}
\figurenum{4}
\epsscale{1.15}
\plotone{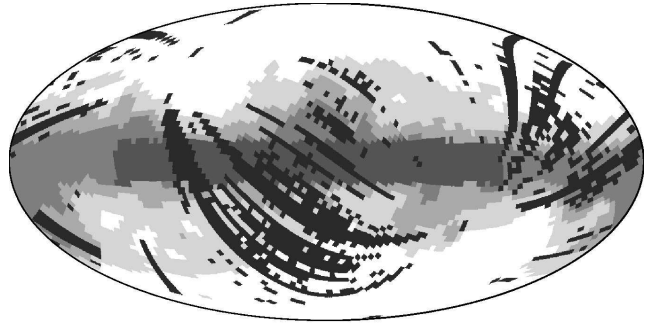}
\caption{A map showing the five nested sky regions for which fits to FIRAS - HFI correlations are made. The FIRAS
pixels shown in black are excluded from all fits due to small numbers of FIRAS observations or contamination by
Jupiter or Mars. The darkest shade of gray shows additional exclusion of the inner Galactic plane, and 
progressively lighter shades show additional exclusion of the Galaxy using HFI 857 GHz brightness cuts of 10, 5, and 2.5 MJy sr$^{-1}$. The sky fractions covered by the five regions are 0.35, 0.53, 0.64, 0.73, and 0.81.}
\end{figure}

For each sky region and each HFI band, we make fits of the form 
\begin{equation}
I^{FIRAS,cc}_{\nu_0} = g I^{HFI}_{\nu_0} + b
\end{equation}
where $I^{FIRAS,cc}_{\nu_0}$ is the color-corrected FIRAS data, $g$ is the relative gain,
and $b$ is the FIRAS intercept.
The fits minimize $\chi^2$ calculated using the inverse noise covariance matrix for the band 
averaged FIRAS data. The calculation of the covariance matrix is described in Appendix B.  
In comparison to the FIRAS uncertainties, our estimated uncertainties for the smoothed HFI data 
are negligible. This is why $I^{HFI}_{\nu_0}$ is treated as the independent variable.

\begin{deluxetable*}{ccccccc}
\tiny
\vspace{-1.0cm}
\tablewidth{0pt}
\tablecaption{FIRAS-HFI Fit Results}
\tablehead{
\colhead{HFI Band} &
\colhead{Sky Region} &
\colhead{Fit Type\tablenotemark{a}}&
\colhead{$g$} &
\colhead{$b$} &
\colhead{$\chi^2_{\nu}$}&
\colhead{$\nu$}\\
\colhead{(GHz)}&
\colhead{} &
\colhead{}&
\colhead{} &
\colhead{(MJy sr$^{-1}$)} &
\colhead{}&
\colhead{}
}
\startdata
100 &  1 &  1  & $  1.6273\pm1.1229 $ & $  0.001\pm0.013 $ &  1.09 &  2121 \\
100 &  1 &  2  & $  1.0000\pm0.0009 $ & $  0.004\pm0.008 $ &  1.09 &  2122 \\
100 &  2 &  1  & $  0.3465\pm0.8441 $ & $  0.009\pm0.013 $ &  1.08 &  3274 \\
100 &  2 &  2  & $  1.0000\pm0.0009 $ & $  0.004\pm0.008 $ &  1.08 &  3275 \\
100 &  3 &  1  & $  1.2014\pm0.6810 $ & $  0.000\pm0.015 $ &  1.10 &  3919 \\
100 &  3 &  2  & $  1.0000\pm0.0009 $ & $  0.002\pm0.009 $ &  1.10 &  3920 \\
100 &  4 &  1  & $  1.2217\pm0.3859 $ & $ -0.004\pm0.016 $ &  1.09 &  4455 \\
100 &  4 &  2  & $  1.0000\pm0.0009 $ & $ -0.001\pm0.010 $ &  1.09 &  4456 \\
100 &  5 &  1  & $  1.0482\pm0.1343 $ & $ -0.010\pm0.021 $ &  1.10 &  4944 \\
100 &  5 &  2  & $  1.0000\pm0.0009 $ & $ -0.009\pm0.013 $ &  1.10 &  4945 \\
    &    &     &                      &                    &       &       \\
143 &  1 &  1  & $  0.3390\pm0.4936 $ & $  0.010\pm0.009 $ &  1.03 &  2121 \\
143 &  1 &  2  & $  1.0000\pm0.0007 $ & $  0.000\pm0.006 $ &  1.03 &  2122 \\
143 &  2 &  1  & $  0.6753\pm0.3599 $ & $  0.009\pm0.009 $ &  1.03 &  3274 \\
143 &  2 &  2  & $  1.0000\pm0.0007 $ & $  0.003\pm0.006 $ &  1.03 &  3275 \\
143 &  3 &  1  & $  0.9627\pm0.2647 $ & $  0.004\pm0.010 $ &  1.05 &  3919 \\
143 &  3 &  2  & $  1.0000\pm0.0007 $ & $  0.003\pm0.006 $ &  1.05 &  3920 \\
143 &  4 &  1  & $  0.9450\pm0.1278 $ & $  0.001\pm0.011 $ &  1.06 &  4455 \\
143 &  4 &  2  & $  1.0000\pm0.0007 $ & $ -0.001\pm0.007 $ &  1.06 &  4456 \\
143 &  5 &  1  & $  0.9499\pm0.0488 $ & $ -0.001\pm0.014 $ &  1.08 &  4944 \\
143 &  5 &  2  & $  1.0000\pm0.0007 $ & $ -0.004\pm0.009 $ &  1.08 &  4945 \\
    &    &     &                      &                    &       &       \\
217 &  1 &  1  & $  0.8673\pm0.1938 $ & $  0.034\pm0.011 $ &  0.79 &  2121 \\
217 &  1 &  2  & $  1.0000\pm0.0016 $ & $  0.026\pm0.003 $ &  0.79 &  2122 \\
217 &  2 &  1  & $  1.2028\pm0.1023 $ & $  0.017\pm0.008 $ &  0.81 &  3274 \\
217 &  2 &  2  & $  1.0000\pm0.0016 $ & $  0.032\pm0.003 $ &  0.81 &  3275 \\
217 &  3 &  1  & $  1.0320\pm0.0521 $ & $  0.028\pm0.005 $ &  0.82 &  3919 \\
217 &  3 &  2  & $  1.0000\pm0.0016 $ & $  0.031\pm0.004 $ &  0.82 &  3920 \\
217 &  4 &  1  & $  1.0083\pm0.0165 $ & $  0.030\pm0.005 $ &  0.81 &  4455 \\
217 &  4 &  2  & $  1.0000\pm0.0016 $ & $  0.031\pm0.004 $ &  0.81 &  4456 \\
217 &  5 &  1  & $  0.9872\pm0.0053 $ & $  0.033\pm0.007 $ &  0.82 &  4944 \\
217 &  5 &  2  & $  1.0000\pm0.0016 $ & $  0.030\pm0.005 $ &  0.82 &  4945 \\
    &    &     &                      &                    &       &       \\
353 &  1 &  1  & $  0.9672\pm0.1132 $ & $  0.039\pm0.026 $ &  0.90 &  2121 \\
353 &  1 &  1e & $  0.9946\pm0.1208 $ & $  0.031\pm0.027 $ &  0.90 &  2121 \\
353 &  1 &  2  & $  1.0000\pm0.0078 $ & $  0.032\pm0.005 $ &  0.90 &  2122 \\
353 &  2 &  1  & $  1.0796\pm0.0385 $ & $  0.014\pm0.012 $ &  0.91 &  3274 \\
353 &  2 &  1e & $  1.0767\pm0.0410 $ & $  0.015\pm0.013 $ &  0.91 &  3274 \\
353 &  2 &  2  & $  1.0000\pm0.0078 $ & $  0.037\pm0.005 $ &  0.91 &  3275 \\
353 &  3 &  1  & $  1.0252\pm0.0172 $ & $  0.026\pm0.009 $ &  0.91 &  3919 \\
353 &  3 &  1e & $  1.0203\pm0.0183 $ & $  0.030\pm0.009 $ &  0.91 &  3919 \\
353 &  3 &  2  & $  1.0000\pm0.0078 $ & $  0.036\pm0.006 $ &  0.91 &  3920 \\
353 &  4 &  1  & $  0.9999\pm0.0045 $ & $  0.036\pm0.009 $ &  0.90 &  4455 \\
353 &  4 &  1e & $  0.9970\pm0.0048 $ & $  0.039\pm0.010 $ &  0.90 &  4455 \\
353 &  4 &  2  & $  1.0000\pm0.0078 $ & $  0.036\pm0.009 $ &  0.90 &  4456 \\
353 &  5 &  1  & $  0.9900\pm0.0038 $ & $  0.045\pm0.011 $ &  1.00 &  4944 \\
353 &  5 &  1e & $  0.9920\pm0.0040 $ & $  0.047\pm0.012 $ &  0.99 &  4944 \\
353 &  5 &  2  & $  1.0000\pm0.0078 $ & $  0.036\pm0.010 $ &  1.00 &  4945 \\
    &    &     &                      &                    &       &       \\
545 &  1 &  1  & $  1.0960\pm0.0673 $ & $ -0.028\pm0.047 $ &  0.80 &  2121 \\
545 &  1 &  1e & $  1.0636\pm0.0481 $ & $  0.009\pm0.033 $ &  0.88 &  2121 \\
545 &  1 &  2e & $  1.0085\pm0.0035 $ & $  0.046\pm0.007 $ &  0.88 &  2122 \\
545 &  2 &  1  & $  1.0429\pm0.0221 $ & $  0.008\pm0.025 $ &  0.82 &  3274 \\
545 &  2 &  1e & $  1.0331\pm0.0153 $ & $  0.031\pm0.017 $ &  0.84 &  3274 \\
545 &  2 &  2e & $  1.0085\pm0.0035 $ & $  0.054\pm0.008 $ &  0.84 &  3275 \\
545 &  3 &  1  & $  1.0348\pm0.0080 $ & $  0.013\pm0.022 $ &  0.82 &  3919 \\
545 &  3 &  1e & $  1.0239\pm0.0055 $ & $  0.038\pm0.015 $ &  0.83 &  3919 \\
545 &  3 &  2e & $  1.0085\pm0.0035 $ & $  0.059\pm0.010 $ &  0.83 &  3920 \\
545 &  4 &  1  & $  1.0173\pm0.0032 $ & $  0.033\pm0.020 $ &  0.82 &  4455 \\
545 &  4 &  1e & $  1.0104\pm0.0022 $ & $  0.052\pm0.014 $ &  0.82 &  4455 \\
545 &  4 &  2e & $  1.0085\pm0.0035 $ & $  0.056\pm0.013 $ &  0.82 &  4456 \\
545 &  5 &  1  & $  1.0155\pm0.0027 $ & $  0.042\pm0.025 $ &  0.97 &  4944 \\
545 &  5 &  1e & $  1.0067\pm0.0020 $ & $  0.061\pm0.019 $ &  1.13 &  4944 \\
545 &  5 &  2e & $  1.0085\pm0.0035 $ & $  0.055\pm0.015 $ &  1.13 &  4945 \\
    &    &     &                      &                    &       &       \\
857 &  1 &  1  & $  1.0429\pm0.0115 $ & $  0.006\pm0.020 $ &  1.12 &  2121 \\
857 &  1 &  2  & $  1.0187\pm0.0025 $ & $  0.043\pm0.006 $ &  1.12 &  2122 \\
857 &  2 &  1  & $  1.0258\pm0.0033 $ & $  0.032\pm0.012 $ &  1.15 &  3274 \\
857 &  2 &  2  & $  1.0187\pm0.0025 $ & $  0.048\pm0.007 $ &  1.15 &  3275 \\
857 &  3 &  1  & $  1.0205\pm0.0015 $ & $  0.045\pm0.011 $ &  1.24 &  3919 \\
857 &  3 &  2  & $  1.0187\pm0.0025 $ & $  0.051\pm0.010 $ &  1.24 &  3920 \\
857 &  4 &  1  & $  1.0170\pm0.0007 $ & $  0.053\pm0.011 $ &  1.45 &  4455 \\
857 &  4 &  2  & $  1.0187\pm0.0025 $ & $  0.043\pm0.017 $ &  1.45 &  4456 \\
857 &  5 &  1  & $  1.0195\pm0.0014 $ & $  0.032\pm0.042 $ & 14.87 &  4944 \\
857 &  5 &  2  & $  1.0187\pm0.0025 $ & $  0.038\pm0.022 $ & 14.87 &  4945
\enddata
\tablenotetext{a}{FIRAS = $g$\thinspace HFI + $b$. Fit type 1: $g$ and $b$ are both fit parameters. Fit type 2: $g$ is fixed and $b$ is a fit parameter. Fit types 1e and 2e indicate fits that use
FIRAS channel-averaged data with questionable FIRAS channels excluded.}
\end{deluxetable*}

\begin{figure} [ht]
\figurenum{5}
\includegraphics[ height=7.6cm, width=8.8cm]{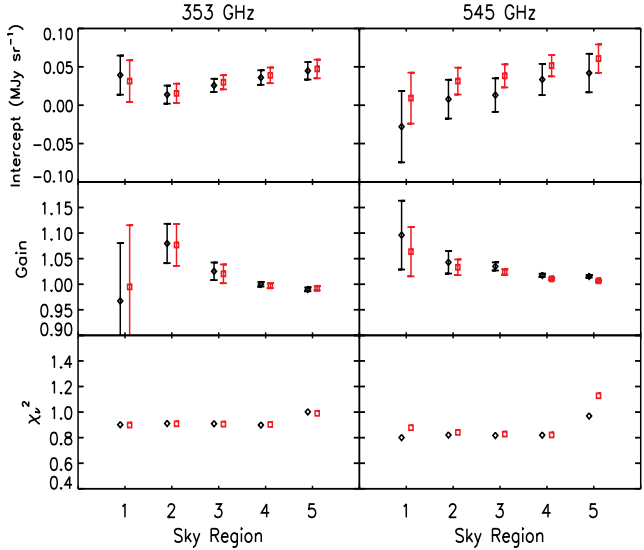}
\caption{Comparison of FIRAS - HFI fit results for cases where questionable FIRAS channels are 
either included (black diamonds) or excluded (red squares) in forming the channel-averaged FIRAS data.
The different symbols are offset from each other for visibility. For the 353 GHz band,
there is no 
significant difference in the results and we adopt those for the case where all of the channels
are included. For the 545 GHz band,
the higher gain values for the case where the questionable channels are included may be
due to changes in dichroic filter frequency response across the channels, so we adopt
results for the case where the questionable channels are excluded. Our final fit 
results for all of the HFI bands are shown in Figure 8.}
\end{figure}

\begin{figure*} [hb]
\figurenum{6}
\epsscale{1.23}
\plotone{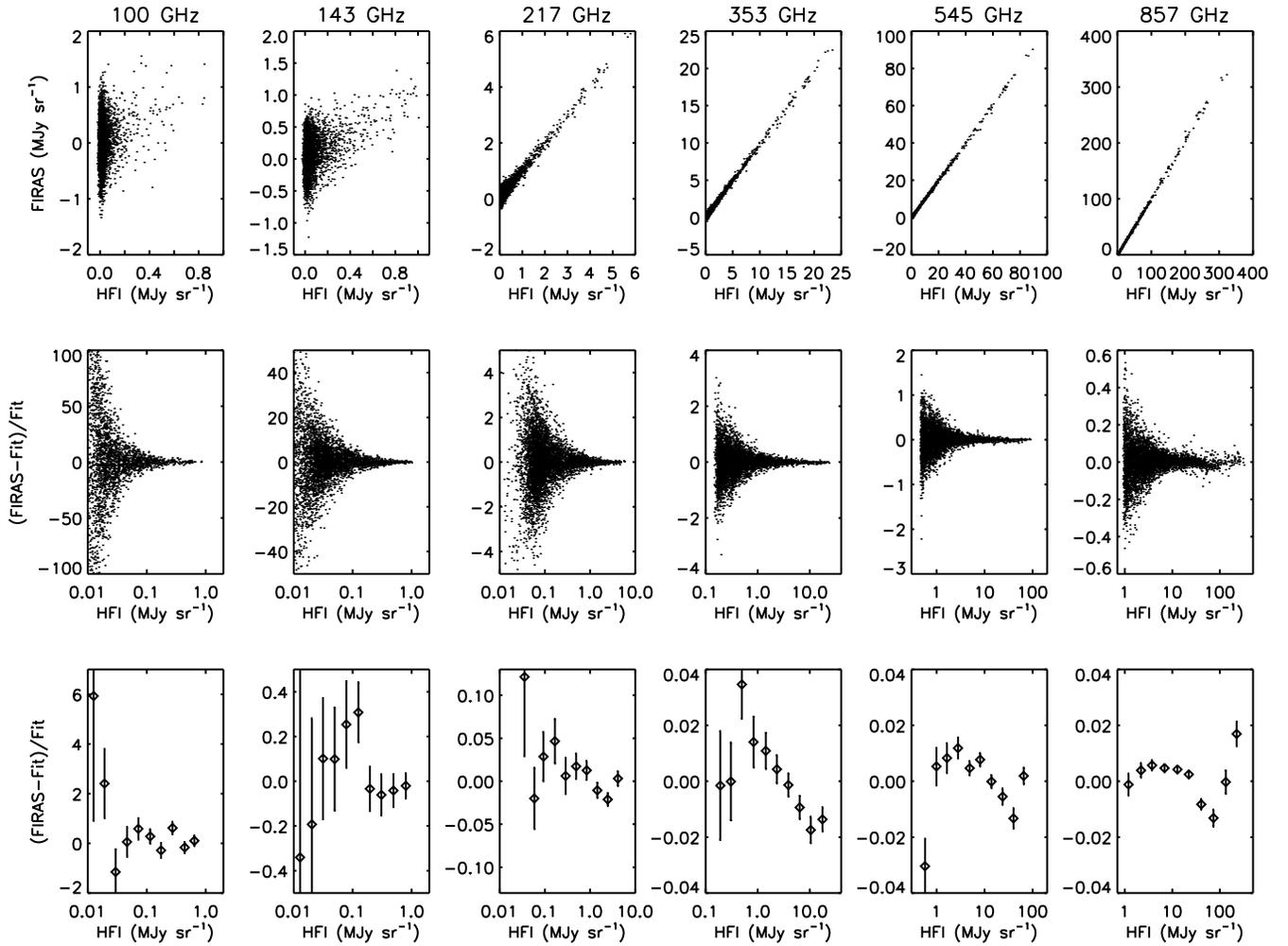}
\caption{The top panels show the FIRAS-HFI correlation for sky region 5 for each HFI band,
the middle panels show the fractional residual of the fit with fixed gain as a
function of HFI brightness, and the bottom panels show the mean fractional residual
and its standard error for logarithmic bins in HFI brightness. Deviations from linearity
are small, at most about 2\% in the 353 to 857 GHz bands.}
\end{figure*}

The results of the fitting are given in Table 4. We made an initial set of fits with both
$g$ and $b$ treated as free parameters. The parameter uncertainties for these fits 
were determined from the 68\% joint confidence region in parameter space. The uncertainties
listed for $b$ do not include uncertainties in CMB monopole subtraction or zodiacal light subtraction.

For the 353 GHz 
and 545 GHz bands, separate fits were made for the cases where questionable FIRAS channels
were either included or excluded in forming the channel-averaged FIRAS data.  Comparison of 
the fit results for these two cases is shown in Figure 5. 
For the 353 GHz band, no significant difference is seen between the results for the two cases.
Thus, we find no evidence of residual mirror transport mechanism ghosts in the channels between
306 and 333 GHz, and we adopt the fit results for the case where all of the FIRAS
channels were included.  For the 545 GHz band, the FIRAS intercept values for the two cases are 
consistent within the uncertainties but the gain values are larger for the case where all of 
the FIRAS channels are used, by 0.9\% (2.6 $\sigma$) for sky region 5.  This may be a 
result of our neglect of dichroic filter response changes across the channels between 605 and 
687 GHz, and we adopt the fit results for the case where these channels were excluded.
A previous indication of a dichroic filter effect in FIRAS data appears in the differential 
CMB spectrum of Figure 3 of Fixsen (2009).

For the 100 to 353 GHz bands, the values of $g$ from our initial fits are consistent with
unity and their uncertainties are larger than or comparable to the uncertainties of the
2015 HFI gain calibration (0.09\%, 0.07\%, 0.16\%, and 0.78\% at 100, 143, 217, and 353 GHz,
respectively, Planck Collaboration 2015 Results VIII 2016). Thus, comparison with the FIRAS 
data does not place any useful constraints on the HFI gain calibration in these bands 
(but it does provide a check).  
We made a second set of fits for these bands assuming that the HFI and FIRAS gain calibrations
agree within their uncertainties. For each band we made a fit with $g$ fixed at unity, and then
made additional fits with $g$ fixed at unity plus and minus the HFI gain calibration 
uncertainty to assess the effect of this uncertainty on the results. Separate fits to assess
the effect of FIRAS gain uncertainty are not needed. This effect is included in the parameter
uncertainties from our fits since the FIRAS bolometer model gain uncertainty and calibration model
emissivity gain uncertainty are included in the FIRAS noise covariance matrix (see Appendix B).
The uncertainty values listed in Table 4 are the quadrature sum of
the uncertainty from our fit and the contribution of HFI gain uncertainty. 
The second set of fits provide more
precise determinations of $b$ than our initial fits because they are not affected by degeneracy
between $b$ and $g$.

For the 545 and 857 GHz bands, the uncertainties in $g$ from our initial fits are
generally much smaller than the HFI gain calibration uncertainty (6.1\% at 545 GHz 
and 6.4\% at 857 GHz). For these bands, we made a second set of fits with $g$ fixed
at $1.0164 \pm 0.0030$ for 545 GHz and $1.0187 \pm 0.0025$ for 857 GHz. These adopted values
are averages of the results of our initial fits for sky regions 4 and 5 at 545 GHz and
regions 3 and 4 at 857 GHz, and at each frequency the adopted uncertainty encompasses the
uncertainties from our initial fits for the two regions. The initial fit result for region 5
at 857 GHz was not used because of the high $\chi^2$ for this fit. The initial fit results 
for the smaller sky regions were not used because of their larger uncertainties, but they
are consistent with the adopted $g$ values.

\begin{figure*}
\figurenum{7}
\epsscale{1.05}
\plotone{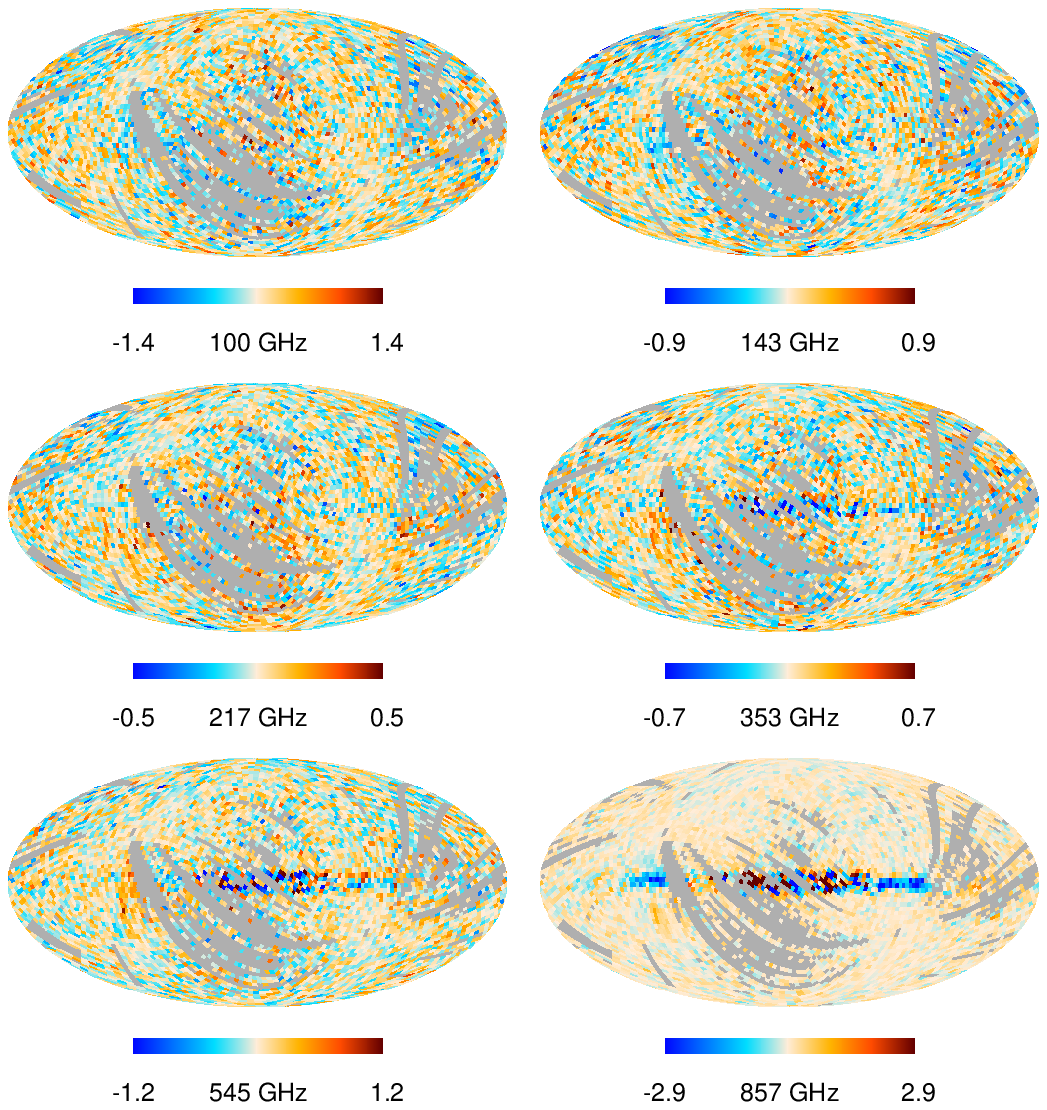}
\caption{Maps of residual brightness (FIRAS minus fit) from the FIRAS-HFI fits for sky region 5
with fixed gain values. The maps are in Galactic coordinates shown in Mollweide projection. 
FIRAS pixels outside of region 5 appear gray. The display ranges shown are in MJy sr$^{-1}$ 
and are $\pm$4 times the rms residual.}
\end{figure*}

\begin{figure*}
\figurenum{8}
\epsscale{1.00}
\plotone{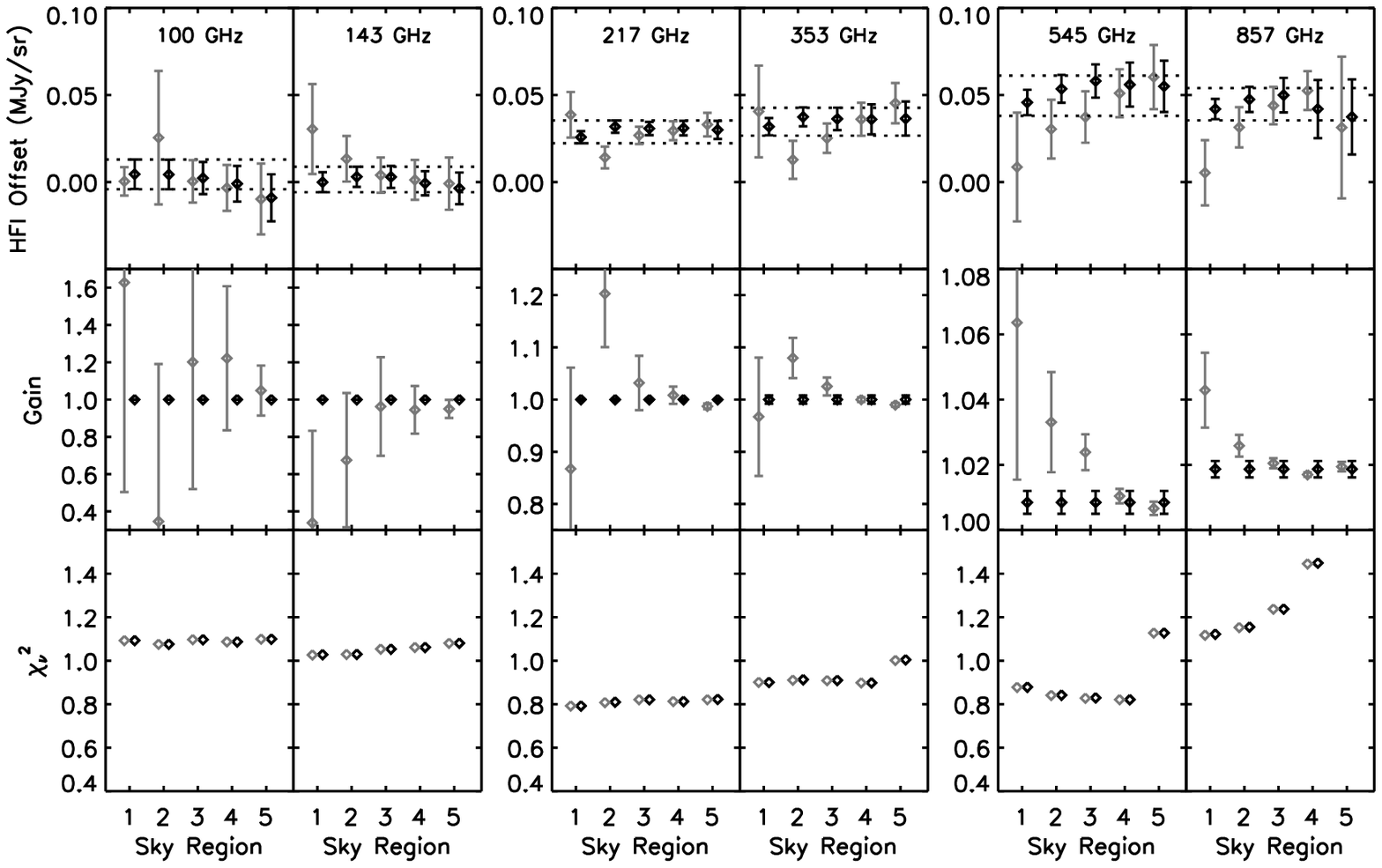}
\caption{Results of FIRAS - HFI fits for different sky regions. The gray symbols show results of
fits for both gain and offset, and black symbols show results of fits with fixed gain. 
The different symbols are offset from each other for visibility, and the plot scale used for the
panels showing the gain changes with frequency.
The top panels show the offset to be added to the HFI map to match the zero level of the FIRAS data,
and the dotted lines show the fit uncertainty range for the value that we adopt for this offset. 
This uncertainty range does not include contributions due to uncertainty in CMB monopole subtraction
or zodiacal light subtraction.}
\end{figure*}

We note that our adopted $g$ results at 545 and 857 GHz are not in conflict with
checks of the 2015 {\it Planck} calibration presented in Planck Collaboration Intermediate 
Results XLVI (2016) 
using measurements of the solar CMB dipole and the first two peaks in the CMB angular
power spectrum in the {\it Planck} bands. For the 545 GHz band they found that a 
calibration based on these CMB measurements is consistent with the 2015 calibration 
within about 1.5\%, and for the 857 GHz band they concluded that the consistency is within 2.5\%.

The degree of linearity between the FIRAS and HFI data is illustrated in Figure 6. The 
figure shows the correlation between FIRAS and HFI data for sky region 5 for each HFI band
and fractional residuals from the fixed $g$ fit for each band as a function of HFI brightness.
Based on the bin-averaged fractional residuals shown in the bottom panels, deviations from 
linearity are small. In the 353 to 857 GHz bands they are at most about 2\%.
Figure 7 shows maps of the fixed $g$ fit residuals for region 5. The most prominent features
appear toward the inner Galactic plane in the high frequency bands. At 857 GHz there is some 
systematic variation with Galactic longitude, with peak-to-peak amplitude of about $\pm5\%$. 
The degree of linearity and the tightness of the correlations that we find are much better
than previously found by Planck Collaboration 2013 Results VIII (2014) using 2013 HFI data and
FIRAS Dust Spectrum Maps (see their Figure C.5).

Fit parameter results for both the initial fits and the fixed $g$ fits are plotted in Figure 8.
Instead of the FIRAS intercept $b$, the top panels of the figure show the HFI offset given 
 by $b/g$. This is the offset to be added to the HFI map to make its zero level consistent
with that of the FIRAS data after CMB and zodiacal light subtraction. For each HFI band, the
HFI offset values from region to region are consistent within
the fit uncertainties. We adopt an average of the offset values from the
fixed gain fits for regions 1 and 2, and we adopt a fit offset uncertainty that encompasses the 
uncertainties from these fits for these regions. 
The adopted fit uncertainty range is shown by the dashed lines in the top panels, and our adopted
HFI offset values and gain values are listed in Table 5. For the HFI offsets, the 
table lists both the uncertainty from the fit and the total uncertainty calculated
as the quadrature sum of fit uncertainty, CMB monopole subtraction uncertainty, and zodiacal light
subtraction uncertainty. We apply the HFI offsets and gains from Table 5 to the {\it Planck} data
release 2.02
zodiacal light subtracted HFI band maps to obtain recalibrated maps that we use for CIB determination.

\begin{deluxetable*}{ccccc}
\scriptsize
\tablewidth{0pt}
\tablecaption{FIRAS\,-\,HFI 2015 Cross-Calibration}
\tablehead{
\colhead{HFI Band} &
\colhead{Gain\tablenotemark{a}} &
\colhead{HFI Offset\tablenotemark{a}} &
\colhead{Offset Uncertainty from Fit} &
\colhead{Total Offset Uncertainty\tablenotemark{b}}\\
\colhead{(GHz)}&
\colhead{} &
\colhead{(MJy sr$^{-1}$)} &
\colhead{(MJy sr$^{-1}$)} &
\colhead{(MJy sr$^{-1}$)}
}
\startdata
100 & 1 & 0.004 & 0.009 & 0.014 \cr
143 & 1 & 0.001 & 0.007 & 0.019 \cr
217 & 1 & 0.029 & 0.007 & 0.023 \cr
353 & 1 & 0.035 & 0.008 & 0.016 \cr
545 & $1.0085 \pm 0.0035$ & 0.050 & 0.012 & 0.015 \cr
857 & $1.0187 \pm 0.0025$ & 0.045 & 0.009 & 0.014
\enddata
\tablenotetext{a}{FIRAS = gain\thinspace(HFI + HFI offset)}
\tablenotetext{b}{Total uncertainty including fit uncertainty,
CMB monopole subtraction uncertainty, and zodiacal light subtraction uncertainty.}
\end{deluxetable*}

The reduced $\chi^2$ values are less than unity for most of our fits for the 217, 353,
and 545 GHz bands, indicating that the FIRAS uncertainties we are using are overestimated in
these bands. We have made test fits with the FIRAS covariance matrix scaled such that the
reduced $\chi^2$ becomes unity. This has negligible effect on the fit parameter values but 
would give a small decrease in fit parameter uncertainties (by about 10\% at 217 GHz).

\section{CIB Determination}

We separate the CIB from Galactic foreground emission in each of the recalibrated, zodiacal light
subtracted HFI band maps using correlations with different template maps that trace the
foreground emission.

CIB values are first obtained by correlating each HFI band map with
Galactic H~I column density for selected sky regions. The Galactic foreground at
these wavelengths is dominated by thermal emission from interstellar dust grains in
equilibrium with the interstellar radiation field. Many studies have shown that Galactic H~I
is a good tracer of this foreground for regions of the diffuse interstellar medium (e.g.,
Boulanger and Perault 1988, Boulanger et al. 1996, Arendt et al. 1998, Planck Collaboration
Intermediate Results XVII 2014) and correlations with H~I have been widely used in previous CIB
determinations at far-infrared to millimeter wavelengths.
The correlations are made for regions where H${_2}$ column density is expected
to be negligible and where intermediate velocity and high velocity gas does not contribute
significantly to the H~I column density, since the emissivity per H atom is lower than normal
for gas at these velocities (e.g., Deul and Burton 1990, Planck Collaboration Early Results XXIV 2011).
We select regions limited to $N_\mathrm{HI} < 3 \times 10^{20}$ cm$^{-2}$, where correlations between 
far-infrared emission and H~I for large sky regions are linear (Boulanger et al. 1996). 
Excess emission relative to $N_\mathrm{HI}$ at higher H~I column densities has been attributed to
emission from dust associated with molecular hydrogen or to nonnegligible optical depth in the 
21-cm line (e.g., Deul and Burton 1992, Reach et al. 1994, Boulanger et al. 1996).  For some
small sky regions correlations have shown excess emission at lower H~I column densities, e.g.,
starting at about $2.0 \times 10^{20}$ cm$^{-2}$ for the Lockman Hole (Arendt et al. 1998).

Correlation with H~I does not take into account any dust emission that may come from the warm ionized
phase of the interstellar medium and is not spatially correlated with H~I. H~II column
density is on average about one-third of H~I column density at high Galactic latitudes (Reynolds 1991),
and the depletion studies of Howk and Savage (1999) and Howk, Sembach, and Savage (2003) suggest that
the dust-to-gas mass ratio in the warm ionized medium is similar to that in the warm neutral medium,
provided that the source of ionization for the warm ionized medium along the lines of sight they studied 
is photoionization by OB stars. These results suggest that
dust emission from the warm ionized medium may be significant. Studies that have correlated high
latitude far infrared or submillimeter observations with a combination of H~I and H$\alpha$ observations,
H~I and pulsar dispersion measures, or H~I and [CII] 158 $\mu$m observations have found CIB values that
are not significantly different from CIB values obtained from correlation using H~I only (Arendt et al. 1998,
Fixsen et al. 1998, Lagache et al. 2000, Odegard et al. 2007). However, this may be due to shortcomings of
these observations as tracers of dust emission from the ionized medium. 

Planck Collaboration Early Results XXIV
(2011) and Planck Collaboration Early Results XVIII (2011) presented evidence that any dust emission from the
ionized medium not correlated with H~I is small for regions they studied that have mean H I column density
less than $2 \times 10^{20}$ cm$^{-2}$. Residual maps after correlating {\it Planck} 353, 545, and 857 GHz
data with Green Bank H~I for these regions have angular power spectra that vary as $k^{-1}$, characteristic 
of CIB anisotropies and much flatter than the $k^{-3}$ spectrum of interstellar dust emission. Furthermore,
probability density functions of the residual maps are consistent with a Gaussian distribution and their
widths have very small scatter from field to field, showing no evidence of residual interstellar dust
contamination.  However, from correlations of {\it Planck} 353, 545, and 857 GHz data and IRAS 100 um data
with H~I data for a much larger sky region, Planck Collaboration 2013 Results XI (2014) found excess dust
emission at H~I column densities less than $1 \times 10^{20}$ cm$^{-2}$, which they suggested may be emission
from dust in the ionized medium. We address the effect of this foreground component on CIB determinations
by correlating each HFI band map with a linear combination of H~I and H$\alpha$ data, and we discuss
possible systematic errors due to shortcomings of H$\alpha$ as a tracer of emission from dust in the 
ionized medium.

As a check on our results, we obtain another set of CIB values for the 100 to 545 GHz bands
by correlating the recalibrated map in each of these bands against the recalibrated 857 GHz map
with the CIB value from the H~I and H$\alpha$ correlation analysis subtracted. This 857 GHz map 
traces dust emission from all phases of the interstellar medium, and hence the correlations can be 
made over larger sky regions than the correlations with H~I or H~I and H$\alpha$. A similar analysis
was used by Planck Collaboration 2013 Results VIII (2014) in determining Galactic zero levels for the 
2013 data release HFI maps.

\subsection{HFI-H~I Correlations}

\subsubsection{Data Selection and Preparation}

We use H~I data from the all-sky HI4PI survey (HI4PI Collaboration 2016), which is based on data from
the Effelsberg-Bonn H~I Survey (EBHIS) first data release (Winkel et al. 2016) and the third revision
of the Galactic All-Sky Survey (GASS, Kalberla and Haud 2015). This is currently the best available all-sky H~I
survey, with full spatial sampling of the sky, corrections applied for stray radiation, and angular
resolution of 16.2$\arcmin$. We use the publicly available H I column density map in HEALPix 
format, which the HI4PI collaboration produced by integrating the spectroscopic data over a velocity range
that includes all significant Galactic emission assuming that the line emission is optically thin.
For column density uncertainties, we adopt the estimate of typical EBHIS column density uncertainty
as a function of column density by Winkel et al. (2016) from comparison of independent EBHIS 
observations in overlap regions between different EBHIS fields.  This gives a fractional uncertainty 
of 7.3\% at $N_\mathrm{HI} = 1 \times 10^{20}$ cm$^{-2}$, decreasing to 3.3\% at $N_\mathrm{HI} = 3 \times 10^{20}$ cm$^{-2}$.

We smooth each of the recalibrated HFI band maps to match the resolution of the HI4PI map. For the 100, 143,
217, and 353 GHz maps, we subtract a CMB anisotropy map smoothed in the same way using either the SMICA CMB 
map, the NILC CMB map, or the Commander CMB map from the {\it Planck} team's 2015 data release. For 545 and 857 GHz,
the contribution of CMB anisotropy to the total emission is negligible and we do not subtract it.  We then degrade
the HFI and HI4PI maps to HEALPix $N_\mathrm{side} = 512$ (pixel size $6.9 \arcmin$) using flat weighting, and
use these data to form HFI-H~I correlations for selected sky regions. We estimate measurement uncertainties
for each of these HFI band maps from the variation among individual survey maps that have been smoothed and 
pixelized in the same way, following the method described in \S2.1.

We consider three sky regions with different cuts on H~I column density to exclude regions of significant
H${_2}$ column density or significant H~I optical depth,
$2.0 \times 10^{20}$ cm$^{-2}$, $2.5 \times 10^{20}$ cm$^{-2}$, and $3.0 \times 10^{20}$ cm$^{-2}$. For each
$N_\mathrm{HI}$ cut, we also exclude the following. (1) We exclude regions with significant intermediate velocity
and high velocity gas. Following Wakker (2004), we identify intermediate and high velocity H~I emission for
a given line of sight as emission for which the observed LSR velocity deviates by more than 35 km s$^{-1}$ from
the range of LSR velocity that is allowed according to a simple Galactic rotation model.  We use the HI4PI 
spectral data to form an H~I column density map for intermediate and high velocity gas defined in this way
and exclude regions where this is greater than 10\% of the total Galactic H~I column density.  (2) We exclude the 
low velocity part of the Magellanic stream using a mask of $10\arcdeg$ radius centered on $l=324.6\arcdeg, 
b=-79.7\arcdeg$
as used by Planck Collaboration Intermediate Results XVII (2014). (3) We exclude pixels within a 16.2$\arcmin$ radius  
of bright point sources, using source positions and fluxes from the second {\it Planck} Catalog of Compact Sources 
and flux
thresholds of 120, 170, 250, 400, 600, and 1000 mJy at 100, 143, 217, 353, 545, and 857 GHz, respectively.
The regions that we use for 857 GHz - H~I correlations are shown in panel (a) of Figure 9.

\begin{figure}
\figurenum{9}
\epsscale{1.00}
\plotone{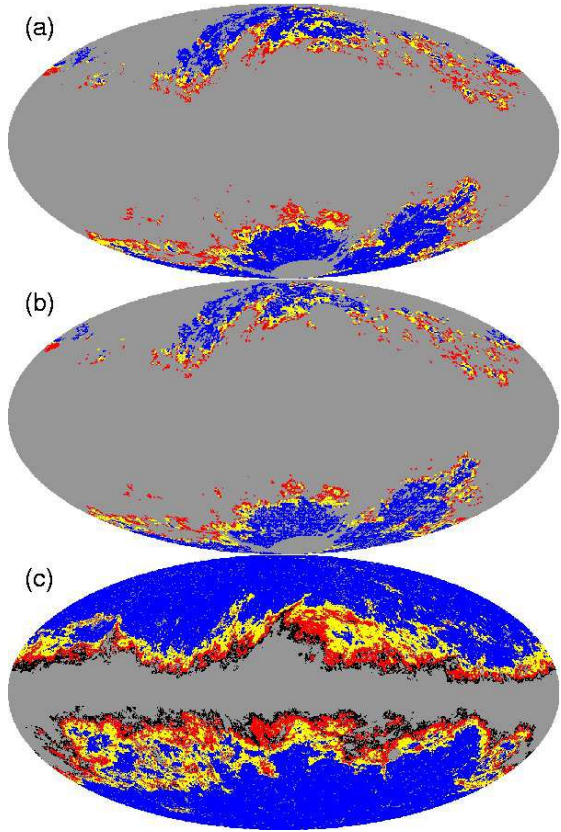}
\caption{Maps showing sky regions used in our correlation analyses. (a) Three 
nested regions for which fits to HFI 857 GHz - H~I correlations are made. The regions for 
cuts on H~I column density of $2.0 \times 10^{20}$ cm$^{-2}$, 
$2.5 \times 10^{20}$ cm$^{-2}$, and $3.0 \times 10^{20}$ cm$^{-2}$ are shown as blue, blue
plus yellow, and blue plus yellow plus red, respectively. 
The sky fractions covered by the three regions are 
0.10, 0.14, and 0.18.
(b) Three regions for which HFI 857 GHz data are fit to a linear
combination of H~I and H$\alpha$ data, for the same H~I cuts with additional
data exclusion based on the H$\alpha$ data (see text).
The sky fractions covered by the three regions are 0.07, 0.10, and 0.13. 
(c) Four nested regions for which 
HFI 545 GHz data are fit with the 857 GHz template. The regions are determined by four different 
cuts on 857 GHz brightness.  The sky fraction ranges from
$\sim$0.41 for the region shown as blue to $\sim$0.68 for the region shown as blue plus yellow
plus red plus black. 
The sky regions used for other HFI bands are similar since only the HFI point source masking changes.
\label{fig:masks857}
}
\end{figure}

\subsubsection{Analysis}

For each sky region and each HFI band, we make a linear fit of the form
\begin{equation}
I_\mathrm{HFI} = a_1 N_\mathrm{HI} + c_1
\end{equation}
where $I_\mathrm{HFI}$ is the smoothed recalibrated HFI data with zodiacal light and CMB anisotropies subtracted.
The fits 
are made using a procedure that minimizes $\chi^2$ calculated using uncertainties in both variables 
(Press et al. 1992). We use an iterative procedure similar to that described by Arendt 
et al. (1998), in which the $N_\mathrm{HI}$ cut is applied using a cut line that is perpendicular to the fit line 
on a plot where each variable is divided by its uncertainty. Planck Collaboration Intermediate Results XVII (2014)
have shown that the scatter in high latitude HFI 857 GHz - H~I correlations is dominated by CIB fluctuations
and variations of dust emissivity per neutral H atom, so for uncertainties in $I_\mathrm{HFI}$ we use 
\begin{equation}
\sigma_\mathrm{HFI,p} = (\sigma_\mathrm{meas,p}^2 + \sigma_\mathrm{CIB}^2 + \sigma_\mathrm{inflate}^2)^{0.5}
\end{equation}
where p is a pixel index, $\sigma_\mathrm{meas,p}$ is our HFI measurement uncertainty estimate for each pixel,
$\sigma_\mathrm{CIB}$ is the rms CIB fluctuation estimated from CIB anisotropy map simulations,
and $\sigma_\mathrm{inflate}$ is a noise inflation term that we adjust such that the fit gives a 
value of unity for $\chi^2$ per degree of freedom.  

The CIB map simulations are based on 
CIB power spectrum measurements from Planck Collaboration 2013 Results XXX (2014). We used their CIB 
bandpowers for multipole bins covering the range $150 < l < 787$ for the 143, 271, 353, 545, and
857 GHz bands. (Measurements are not available for 100 GHz and we set $\sigma_{CIB}$ to zero for this band.)
For each frequency, we fit the measured power spectrum with the sum of a halo model spectral template
from Mak et al. (2017) plus a constant to obtain a power spectrum for individual $l$ values
from 2 to 1160. The spectral templates are available at 353, 545, and 857 GHz and vary only slightly 
with frequency. We used the 353 GHz template for the fits to the 143 and 217 GHz power spectra.
The spectral template fits were used to generate simulated CIB anisotropy maps with 16.2$\arcmin$ resolution FWHM.

The $\sigma_\mathrm{CIB}$ values obtained from these maps are listed in Table 6, together with mean 
$\sigma_\mathrm{meas}$ values and $\sigma_\mathrm{inflate}$ values for the $2.5 \times 10^{20}$ cm$^{-2}$ 
$N_\mathrm{HI}$ cut sky region. For each of the 100 to 353 GHz bands, the $\sigma_\mathrm{inflate}$ value
listed is an average of the values used for the different cases of CMB map subtraction.  The spectrum of 
the $\sigma_\mathrm{inflate}$ values is somewhat flatter than a characteristic interstellar dust spectrum, 
so in the lower frequency bands $\sigma_\mathrm{inflate}$ probably accounts for residual CMB fluctuations
in addition to dust emissivity variations.
 
We obtain fit parameter uncertainties from the 68\% joint confidence region in parameter space,
but for the fit intercepts we adopt more conservative uncertainties based on the variation among 
intercept values from separate fits made for subsets of the selected sky region. We divide each sky 
region into six subregions defined by
$0\arcdeg < l < 120\arcdeg$, $120\arcdeg < l < 240\arcdeg$, and $240\arcdeg < l < 360\arcdeg$ for 
$b < 0\arcdeg$ and the same longitude ranges for $b > 0\arcdeg$. We adopt a
weighted standard deviation of intercept values for the six subregions as our uncertainty
estimate. This is calculated as the square root of the unbiased estimator of weighted variance
for weights given by 1/$\sigma_i^2$, where $\sigma_i$ is the intercept fit uncertainty for 
subregion $i$.

\subsubsection{Results}

\begin{figure*}
\figurenum{10}
\epsscale{1.00}
\plotone{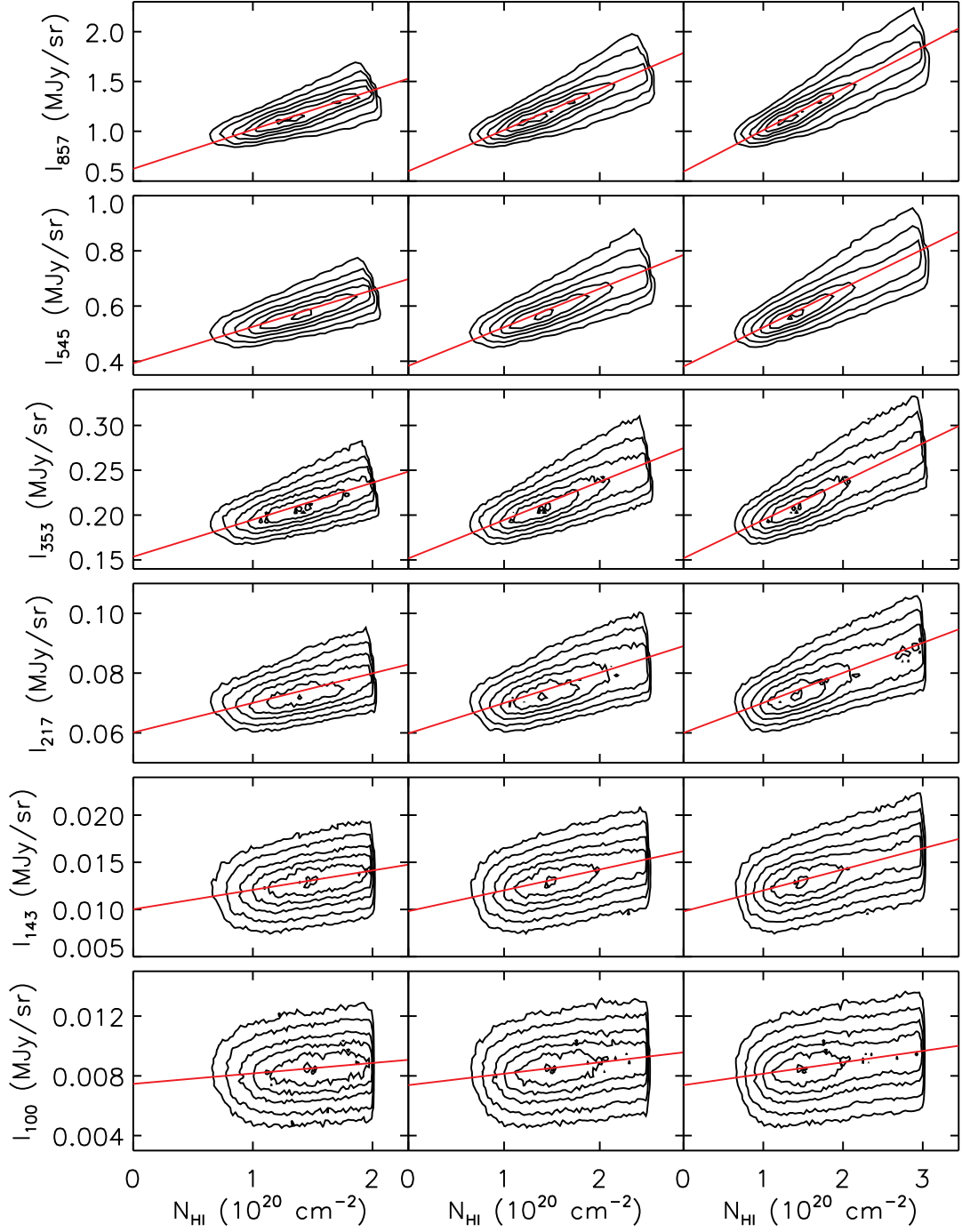}
\caption{Linear fits to correlations between recalibrated, zodiacal light 
subtracted HFI data and H~I column density for cuts on $N$(H~I) of
$2.0 \times 10^{20}$ cm$^{-2}$ (left panels), $2.5 \times 10^{20}$ cm$^{-2}$ 
(center panels), and $3.0 \times 10^{20}$ cm$^{-2}$ (right panels). 
The {\it Planck} SMICA CMB map was subtracted from the HFI data for the 100, 143, 217,
and 353 GHz bands. The contours
show the density of points for individual pixels, and the contour levels are 
5\%, 15\%, 35\%, 55\%, 75\%, and 95\% of the peak density for each panel.
}
\end{figure*}

Figure 10 shows our linear fits to the HFI - H~I correlations for the three sky regions, for the case
where the SMICA CMB map was subtracted from the 100, 143, 217, and 353 GHz data. 
The correlations for NILC CMB subtraction appear similar; those for Commander CMB subtraction
are somewhat tighter at the lowest frequencies.  None of the correlations show any 
significant excess dust emission relative to the fit at the higher column densities, so any contribution 
from H${_2}$ associated dust or any effect of 21-cm line optical depth appears to be small. 

\begin{figure*}
\figurenum{11}
\epsscale{1.2}
\plotone{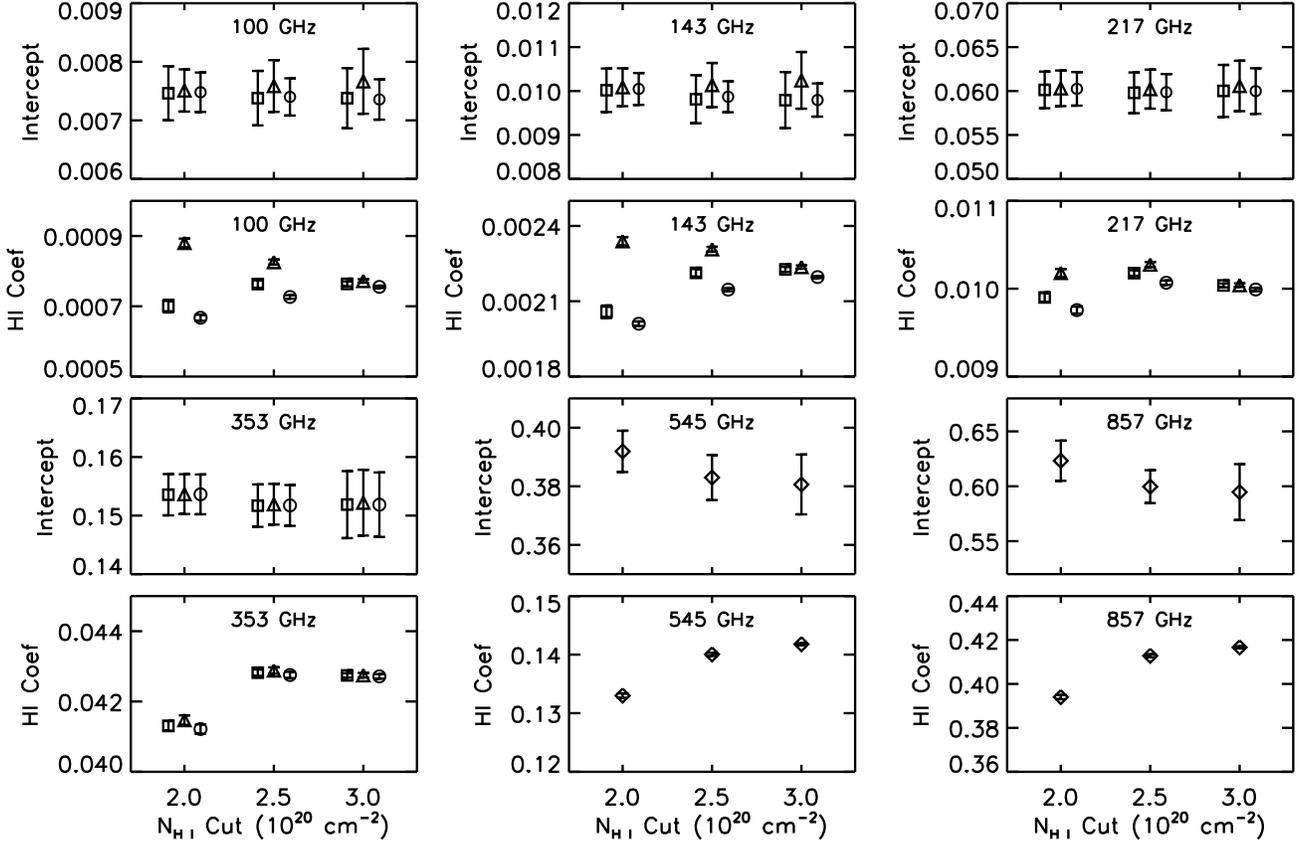}
\caption{Parameters from linear fits to HFI-HI correlations as a function
of the cut on H~I column density used to determine the sky region.
The intercept values are in MJy sr$^{-1}$ and the slope values are
in MJy sr$^{-1}$/10$^{20}$ cm$^{-2}$.
For the 100 to 353 GHz bands, results are shown for subtraction of the
SMICA CMB map (squares), the NILC CMB map (triangles), and the Commander 
CMB map (circles) from the HFI data, with the different plot symbols offset 
from each other for visibility. The intercept results are consistent within 
the adopted uncertainties for the different cuts and for 
the different CMB map subtractions for each cut.}
\end{figure*}

Figure 11 shows the fit parameters as a function of $N_\mathrm{HI}$ cut.
The intercept results are consistent within the adopted uncertainties for the different cuts, and for 
the different CMB subtractions for each cut. In the lower frequency bands, the HI coefficients for 
the different CMB subtractions vary by more than the fit uncertainties. 
We adopt the results for the $2.5 \times 10^{20}$ cm$^{-2}$ cut.  For the
100 to 353 GHz bands we adopt straight averages of the intercept values, intercept uncertainties,
and HI coefficients for the different CMB subtractions, and we adopt increased HI coefficient 
uncertainties that encompass the fit results for the different CMB subtractions.
These adopted fit parameters are listed in Table 6.

\begin{deluxetable*}{cccccc}
\scriptsize
\tablewidth{0pt}
\tablecaption{HFI vs. H~I Fit Parameters\tablenotemark{a}}
\tablehead{
\colhead{HFI Band} &
\colhead{$\langle \sigma_\mathrm{meas} \rangle$\tablenotemark{b}} &
\colhead{$\sigma_\mathrm{CIB}$\tablenotemark{c}} &
\colhead{$\sigma_\mathrm{inflate}$\tablenotemark{d}} &
\colhead{Slope $a_1$} &
\colhead{Intercept $c_1$}\\
\colhead{(GHz)} &
\colhead{(MJy sr$^{-1}$)} &
\colhead{(MJy sr$^{-1}$)} &
\colhead{(MJy sr$^{-1}$)} &
\colhead{(MJy sr$^{-1}$/10$^{20}$ cm$^{-2}$)} &
\colhead{(MJy sr$^{-1}$)}
}
\startdata
 100 &  0.0007  &   0        &  0.0014  & 0.00077  $\pm$ 0.00006  &  0.0075 $\pm$  0.0004 \cr
 143 &  0.0005  &   0.0008   &  0.0018  & 0.00222  $\pm$ 0.00009  &  0.0099 $\pm$  0.0005 \cr
 217 &  0.0010  &   0.0022   &  0.0052  & 0.01017  $\pm$ 0.00013  &  0.060  $\pm$  0.002 \cr
 353 &  0.002   &   0.009    &  0.015   & 0.04282  $\pm$ 0.00015  &  0.152  $\pm$  0.004 \cr
 545 &  0.004   &   0.025    &  0.041   & 0.1400   $\pm$ 0.0003   &  0.383  $\pm$  0.008 \cr
 857 &  0.009   &   0.049    &  0.111   & 0.4128   $\pm$ 0.0007   &  0.600  $\pm$  0.015
\enddata
\tablenotetext{a}{Results for the $2.5 \times 10^{20}$ cm$^{-2}$ $N$(H I) cut sky region.}
\tablenotetext{b}{Mean HFI measurement uncertainty for the sky region.}
\tablenotetext{c}{Estimated rms CIB fluctuation.}
\tablenotetext{d}{Noise inflation term for which the fit gives $\chi_{\nu}^2$ of unity.}
\end{deluxetable*}

\subsection{Correlations with H~I and H$\alpha$}

\subsubsection{Data Selection and Preparation}

We correlate data from each recalibrated HFI band map with a linear combination of H~I and H$\alpha$ 
data. We use H$\alpha$ data from the Wisconsin H-Alpha Mapper (WHAM) all-sky survey first data 
release (Haffner et al. 2003, Haffner et al. 2010, Haffner et al. 2018). The WHAM instrument has a 
$1\arcdeg$ diameter field of view, and the survey was made on a regular Galactic coordinate grid with 
pointings separated by $0.98\arcdeg /$cos $b$ in $l$ and $0.85\arcdeg$ in $b$. We use the velocity 
integrated data provided by the WHAM team. The velocity range used, $-80 < v_\mathrm{LSR} < 80$ km s$^{-1}$, 
includes all significant Galactic emission for the pointings that we use in our analysis.

We apply approximate corrections for extinction and for scattered H$\alpha$ as described by Bennett et al. 
(2013).
The extinction correction assumes that H$\alpha$ emission and extinction are uniformly mixed along each
line of sight. The correction for scattered H$\alpha$ is based on correlations between H$\alpha$ and 
100 $\micron$ emission found by Witt et al. (2010) for selected intermediate latitude dust clouds and 
by Brandt and Draine (2012) 
for Sloan Digital Sky Survey blank sky regions at intermediate to high Galactic latitudes. It assumes
that the spatial variation of the illuminating H$\alpha$ radiation field in the Galaxy is similar to that
of the radiation responsible for dust heating, so that the 100 $\micron$ dust emission map of
Schlegel et al. (1998) can be used as 
a tracer of scattered H$\alpha$. For the pointings used in our analysis, the extinction correction is 
less than 6\%. 
The scattering correction is generally between 0.05 and 0.2 Rayleigh\footnote{1 R = 
$10^{6}/4\pi$ H$\alpha$ photons cm$^{-2}$ s$^{-1}$ sr$^{-1} = 2.4 \times 10^{-7}$ erg cm$^{-2}$ s$^{-1}$ sr$^{-1}$},
and the median scattering correction is 16\%.
 
We use HFI and H~I maps as described in \S3.1. We smooth them to WHAM resolution using a smoothed top-hat
approximation for the WHAM beam response from Finkbeiner (2003), and then sample them at the position
of each WHAM pointing. For the HFI 100, 143, 217, and 353 GHz maps, we subtract a SMICA, NILC, or 
Commander CMB anisotropy map smoothed and sampled in the same way.
We estimate measurement uncertainties for the smoothed HFI data from the variation among
individual survey maps that have been smoothed and sampled in the same way, following the method described
in \S2.1. 
We estimate uncertainties for the smoothed H~I data based on the Winkel et al. (2016) results for EBHIS
column density uncertainty as a function of column density, which they found to be approximately consistent
with a combination of thermal noise and a calibration uncertainty of about 2\%. We subtracted the 
calibration uncertainty component, scaled the residual component to account for smoothing to WHAM 
resolution assuming that it is Gaussian thermal noise, and then added the calibration uncertainty back to
obtain our adopted uncertainty of the smoothed H~I data as a function of column density.  

\begin{figure*}
\figurenum{12}
\epsscale{1.20}
\plotone{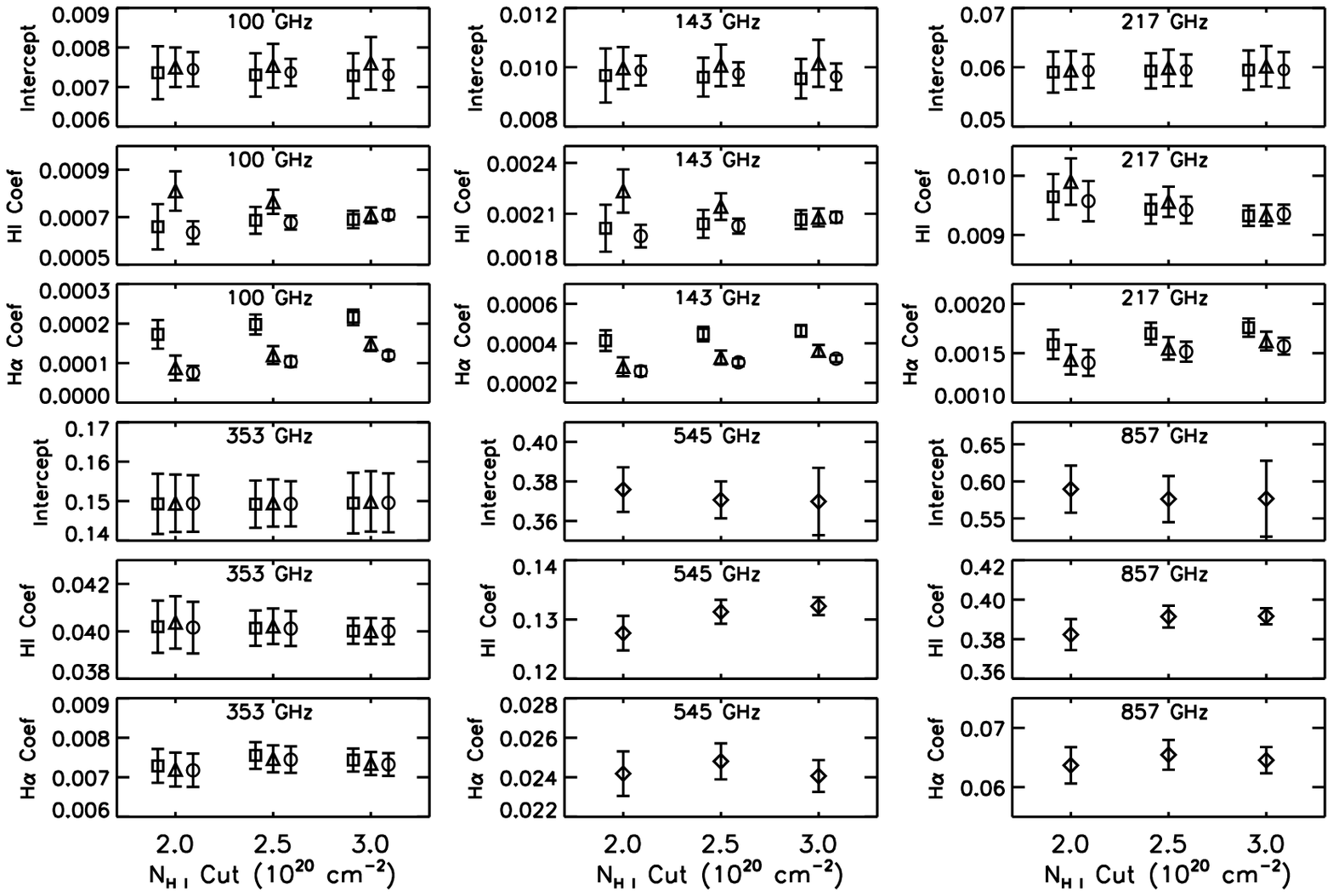}
\caption{Parameters from fits of HFI data to a linear combination of H~I and H$\alpha$ data
as a function of the cut on H~I column density used to determine the sky region.
The intercept values are in MJy sr$^{-1}$, the H~I coefficients are
in MJy sr$^{-1}$/10$^{20}$ cm$^{-2}$, and the H$\alpha$ coefficients are in
MJy sr$^{-1}$/R.
For the 100 to 353 GHz bands, results are shown for subtraction of the
SMICA CMB map (squares), the NILC CMB map (triangles) and the Commander 
CMB map (circles) from the HFI data, with the different plot symbols offset 
from each other for visibility. The intercept results are consistent within the adopted 
uncertainties for the different cuts, and for the different CMB subtractions for each cut.}
\end{figure*}

\begin{figure*}
\figurenum{13}
\epsscale{1.20}
\plotone{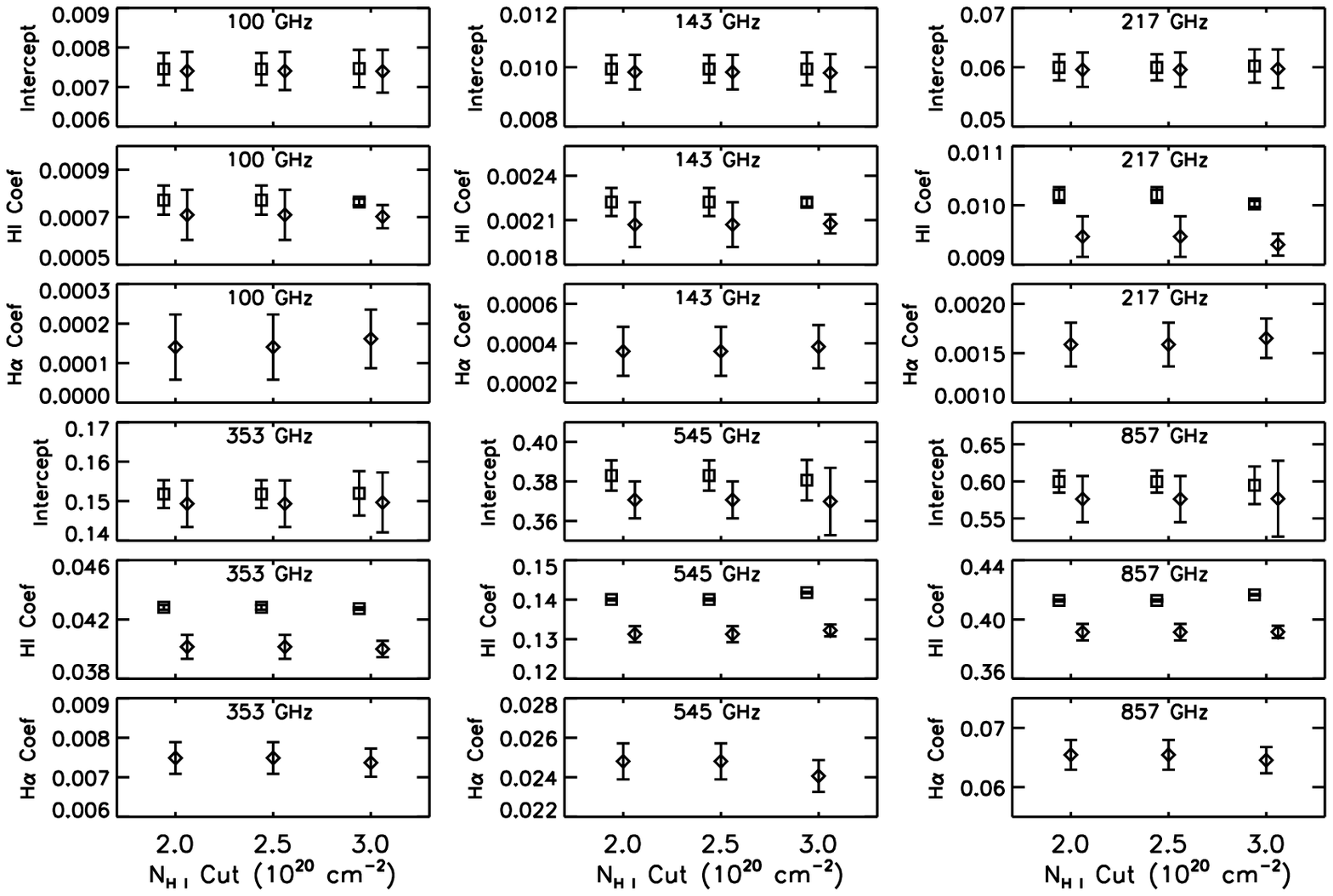}
\caption{Comparison of parameters from fits of HFI data to H~I data (squares) with
those from fits to H~I and H$\alpha$ data (diamonds),
as a function of the cut on H~I column density used to determine
the sky region. 
The intercept values are in MJy sr$^{-1}$, the H~I coefficients are in MJy 
sr$^{-1}$/10$^{20}$ cm$^{-2}$, and the H$\alpha$ coefficients are in MJy sr$^{-1}$/R.
For the 100 to 353 GHz bands, the results shown are averages of results for three
different cases of CMB anisotropy subtraction.  The intercept values for the H~I fits
and the fits including H$\alpha$ are consistent within the uncertainties. The H$\alpha$
correlated component is detected with high significance in the higher frequency bands.}
\end{figure*}

We use three nested sky regions defined by the same cuts on H~I column density as used for the HFI - H~I
correlations. We exclude regions with significant intermediate velocity and high velocity gas
as before using intermediate and high velocity column density maps smoothed to WHAM resolution,
and we exclude the Magellanic stream as before. We exclude WHAM pointings that are within 0.6\arcdeg
of any bright HFI point source, using the same flux thresholds as in \S3.1.1. We exclude pointings
that are flagged as potentially contaminated by a bright star and pointings that were identified
as WHAM point sources by Reynolds et al. (2005). We exclude pointings that sample the H II region
around $\alpha$ Vir (Reynolds 1985) using an exclusion radius of 11$\arcdeg$ centered on $l = 316\arcdeg, 
b = +55\arcdeg$, since this region may not be representative of the general diffuse ionized medium. 
Other faint, high latitude H II regions identified in the WHAM data by Haffner (2001) are excluded by 
the cuts on H~I column density or by the WHAM point source exclusions. The regions that we use for 
the correlations of 857 GHz data with H~I and H$\alpha$ are shown in Figure 9(b).

\subsubsection{Analysis}

For each sky region and each HFI band, we make a linear fit of the form
\begin{equation}
I_\mathrm{HFI} = a_2 N_\mathrm{HI} + b I(H\alpha) + c_2
\end{equation}
where $I_\mathrm{HFI}$, $N_\mathrm{HI}$, and I(H$\alpha$) are the data described above for the selected
WHAM pointings. The fits are made as described in \S3.1, minimizing $\chi^2$ calculated using 
uncertainties in all of the variables. We adopt uncertainties in $I_\mathrm{HFI}$ as given by 
equation (7) using $\sigma_\mathrm{CIB}$ values from the simulated CIB anisotropy maps described in \S3.1
smoothed to WHAM resolution, and with $\sigma_\mathrm{inflate}$ again adjusted to give $\chi^2$ per 
degree of freedom of unity. For the I(H$\alpha$) uncertainties we use the measurement uncertainties
provided in the WHAM data release, which do not include systematic or calibration uncertainties.
Systematic errors associated with removal of geocoronal and atmospheric emission lines from the
WHAM spectra can be larger than the measurement uncertainties but are typically less than 0.1 R.
We find that adding an uncertainty of 0.1 R to the measurement uncertainty for each WHAM pointing 
has negligible effect on the parameter values from the fitting. 

We adopt fit uncertainties from the 68\% joint confidence region in parameter space for the
H~I and H$\alpha$ coefficients. For the intercepts, we adopt more conservative uncertainty
estimates from the variation among intercept values from separate fits for six subregions as described
in \S3.1.

\subsubsection{Results and Discussion}

\begin{figure}
\figurenum{14}
\epsscale{1.10}
\plotone{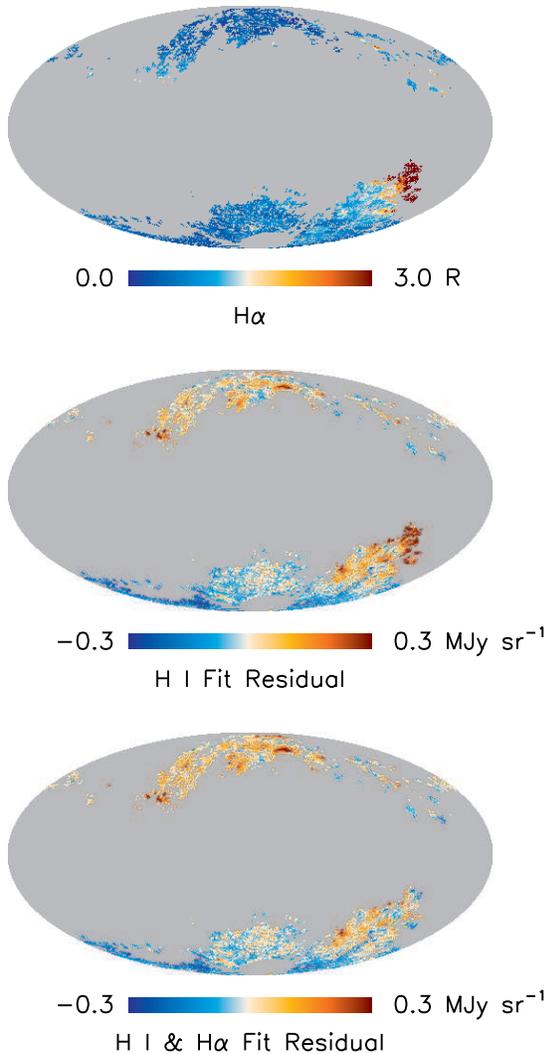}
\caption{Comparison of the H$\alpha$ map corrected for extinction and scattering (top), a map 
of residual 857 GHz emission after an H~I fit is subtracted (middle), and a map of residual 
857 GHz emission after a combined H~I and H$\alpha$ fit is subtracted (bottom), for the sky
region set by the $N_\mathrm{HI}$ cut of $2.5 \times 10^{20}$ cm$^{-2}$.  This shows correlation 
between the H$\alpha$ map and the H~I fit residual map, and improvement in the residual map 
when H$\alpha$ is included in the fit.}
\end{figure}

Figure 12 shows the fit parameters as a function of $N_\mathrm{HI}$ cut.
The intercept results are consistent within the adopted uncertainties for the different cuts, and for 
the different CMB subtractions for each cut. Figure 13 compares the intercepts and H~I coefficients 
with those obtained from the H~I fits in \S3.1. For 100 to 353 GHz the results have been 
averaged over the different CMB subtractions, and we have adopted increased uncertainties for the H~I
and H$\alpha$ coefficients that encompass the fit results for the different CMB subtractions.
The intercepts and H~I coefficients from the fits including H$\alpha$ 
are slightly smaller than those from the H~I fits.
The H$\alpha$ coefficients are significantly greater than zero in the higher frequency bands,
and the shape of the H$\alpha$ coefficient spectrum is consistent with that of the H~I coefficient spectrum.
These results are indications that H$\alpha$ is tracing some dust emission from the ionized phase
of the interstellar medium that is not spatially correlated with H~I, and that the fits including 
H$\alpha$ do a better job of tracing the total Galactic foreground emission than the fits using 
H~I only.

This is shown more directly in Figure 14, which compares the H$\alpha$ map used in our analysis 
for the $N_\mathrm{HI}$ cut of $2.5 \times 10^{20}$ cm$^{-2}$ with
the 857 GHz residual map from the combined H~I and H$\alpha$ fit and an 857 GHz residual map from an
H~I-only fit at WHAM resolution for the same WHAM pointings. It shows a clear correlation between 
the H$\alpha$ map and the residual map from the H~I-only fit, and improvement in the residual map
when H$\alpha$ is included in the fit. The rms residual
is 0.120 MJy sr$^{-1}$ for the H~I-only fit and 0.101 MJy sr$^{-1}$ for the H~I and H$\alpha$ fit.
There is similar fractional improvement for the 353 and 545 GHz bands and smaller fractional improvement
for the lower frequency bands. The improvement in the residuals is largely due to improvement in
the area of highest H$\alpha$ brightness, but for fits with pixels brighter than 3 R excluded the 
857 GHz rms residual still improves from 0.106 MJy sr$^{-1}$ fitting with H~I only to 0.101 MJy sr$^{-1}$ 
fitting with H~I and H$\alpha$.

Based on these comparisons, 
we favor the results from the combined H~I and H$\alpha$ fits over those from
the H~I fits in \S3.1, and adopt the fit intercepts for the $2.5 \times 10^{20}$ cm$^{-2}$ cut as
our preferred CIB values in the HFI bands. The fit parameters for this cut are listed in Table 7.
The uncertainties listed are fit uncertainties. Total CIB uncertainties 
are presented in \S3.4.

Following Lagache et al. (2000) and Odegard et al. (2007), we use a
ratio of $I$(H$\alpha$)/$N$(H$^{+})$ = 1.15 R/$10^{20}$ cm$^{-2}$
to convert the H$\alpha$ coefficient values from our fits to emissivity per H$^{+}$ ion, $\epsilon$(H$^+$).
This is the mean ratio of H$\alpha$ intensity to pulsar dispersion measure found by Reynolds (1991) for
four high latitude lines of sight toward pulsars at $z >$ 4 kpc (greater than the warm ionized medium's
scale height of 0.9-1.8 kpc, e.g., Haffner et al. 1999, Berkhuijsen and M\"uller 2008, 
Gaensler et al. 2008, Savage and Wakker 2009). It corresponds to an effective
electron density, $n_{eff} \equiv \int n_e^2 ds/\int n_e ds$, of 0.08 cm$^{-3}$ for an electron temperature 
of 8000 K, no extinction, and no scattered H$\alpha$.  The $I$(H$\alpha$)/$N$(H$^{+})$ ratio 
ranges from 0.75 to 1.9 R/$10^{20}$ cm$^{-2}$ for the lines of sight studied by Reynolds, 
so the measurement uncertainty for the conversion factor is large.  

\begin{deluxetable*}{ccccccc}
\scriptsize
\tablewidth{0pt}
\tablecaption{HFI vs. H~I and H$\alpha$ Fit Parameters\tablenotemark{a}}
\tablehead{
\colhead{HFI Band} &
\colhead{$\langle \sigma_\mathrm{meas} \rangle$\tablenotemark{b}} &
\colhead{$\sigma_\mathrm{CIB}$\tablenotemark{c}} &
\colhead{$\sigma_\mathrm{inflate}$\tablenotemark{d}} &
\colhead{H~I Coefficient $a_2$} &
\colhead{H$\alpha$ Coefficient $b$} &
\colhead{Intercept $c_2$}\\
\colhead{(GHz)} &
\colhead{(MJy sr$^{-1}$)} &
\colhead{(MJy sr$^{-1}$)} &
\colhead{(MJy sr$^{-1}$)} &
\colhead{(MJy sr$^{-1}$/10$^{20}$ cm$^{-2}$)} &
\colhead{(MJy sr$^{-1}$/R)} &
\colhead{(MJy sr$^{-1}$)}
}
\startdata
 100 &  0.0003  &   0        &  0.0008    & 0.0007   $\pm$ 0.0001  & 0.00014 $\pm$ 0.00008 &  0.0074 $\pm$  0.0005 \cr
 143 &  0.0002  &   0.0004   &  0.0012    & 0.0021   $\pm$ 0.0002  & 0.00036 $\pm$ 0.00012 &  0.0098 $\pm$  0.0006 \cr
 217 &  0.0005  &   0.0013   &  0.0044    & 0.0095   $\pm$ 0.0003   & 0.0016  $\pm$ 0.0002 &  0.060  $\pm$  0.003 \cr
 353 &  0.0010  &   0.0054   &  0.013     & 0.0402   $\pm$ 0.0008   & 0.0075  $\pm$ 0.0004 &  0.149  $\pm$  0.006 \cr
 545 &  0.002   &   0.015    &  0.034     & 0.131    $\pm$ 0.002   &  0.0248  $\pm$ 0.0009 &  0.371  $\pm$  0.009 \cr
 857 &  0.008   &   0.031    &  0.095     & 0.391    $\pm$ 0.006   &  0.0655  $\pm$ 0.0025 &  0.576  $\pm$  0.031
\enddata
\tablenotetext{a}{Results for the $2.5 \times 10^{20}$ cm$^{-2}$ $N$(H I) cut sky region.}
\tablenotetext{b}{Mean HFI measurement uncertainty for the sky region.}
\tablenotetext{c}{Estimated rms CIB fluctuation.}
\tablenotetext{d}{Noise inflation term for which the fit gives $\chi_{\nu}^2$ of unity.}
\end{deluxetable*}

We also note a possible systematic error for the conversion factor. A value higher than the mean 
$I$(H$\alpha$)/$N$(H$^{+})$ ratio (corresponding to a higher $n_{eff}$) may be more appropriate 
if the correlation between dust emission and H$\alpha$ brightness is preferentially determined by regions
of the warm ionized medium with higher than average density, such as the H$\alpha$ emitting regions 
associated with H~I clouds studied by Reynolds et al. (1995) with densities of about 0.2--0.3 cm$^{-3}$.
This would be expected if the abundance of dust decreases with decreasing gas density in the warm ionized
medium, as has been inferred for the neutral atomic medium from interstellar absorption line observations 
(e.g., Jenkins 1987, Savage and Sembach 1996). It has been suggested for the ionized medium by Howk 
and Savage (1999) based on the variation of aluminum depletion with mean electron density they found for
two lines of sight that sample the warm ionized medium and three lines of sight that sample low density 
H II regions around early type stars.

\begin{figure}
\figurenum{15}
\epsscale{1.20}
\plotone{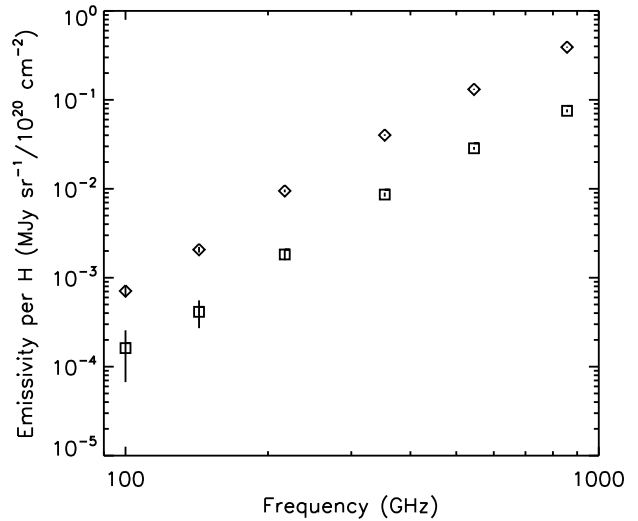}
\caption{Comparison of dust emissivity per H atom for the neutral atomic gas phase (diamonds) and the
dust emissivity per H ion for the warm ionized gas phase (squares). The emissivity per H$^{0}$
values are the H~I coefficients from the combined H~I and H$\alpha$ fits. The
emissivity per H$^{+}$ values are obtained from the H$\alpha$ coefficients using a
conversion factor of $I(H\alpha$)/$N(H^{+})$ = 1.15 R/$10^{20}$ cm$^{-2}$. The error bars
show fit uncertainties and do not include uncertainty in the conversion factor.
}
\end{figure}

Figure 15 shows a comparison of the spectrum of the emissivity per H$^{+}$ ion using this conversion factor
with the spectrum of emissivity per neutral H atom $\epsilon$(H$^0$) as given by the H~I coefficients $a_2$ from 
our fits.  The spectral shapes are consistent, so there is no evidence for differences in the 
mean radiation field or mean dust optical properties between the neutral and ionized gas phases. The emissivity
per H$^{+}$ values are about 20\% of the emissivity per H$^{0}$ values.

This result is consistent
with the upper limit of 40\% for the 100 $\micron$ $\epsilon$(H$^+$)/$\epsilon$(H$^0$) ratio found by Odegard 
et al. (2007) from analysis of DIRBE, WHAM H$\alpha$, and H~I observations for a
much smaller sky region. They
discussed possible explanations for their low derived $\epsilon$(H$^+$)/$\epsilon$(H$^0$) ratio.
Available evidence suggested that it is at least partly due to a lower dust-to-gas mass ratio in the ionized
medium than in the neutral medium and partly due to a shortcoming of using H$\alpha$ as a tracer of emission 
from dust in the 
ionized medium. Error in the adopted $I$(H$\alpha$)/$N$(H$^{+})$ conversion factor could also contribute.
Another possible factor is that the H~I fit coefficient may overestimate $\epsilon$(H$^0$) somewhat if there is a 
component of dust emission from the ionized medium that is correlated with $N$(H~I).
If H$\alpha$ is not a good tracer, they showed that a correlation analysis like ours using H~I and 
H$\alpha$ would be expected to underestimate the emissivity per H$^{+}$ ion and overestimate the CIB.
Appendix C presents an estimate of the possible effect on our CIB results, which suggests that it is 
smaller than our adopted CIB uncertainties.

\begin{figure*}
\figurenum{16}
\epsscale{0.7}
\plotone{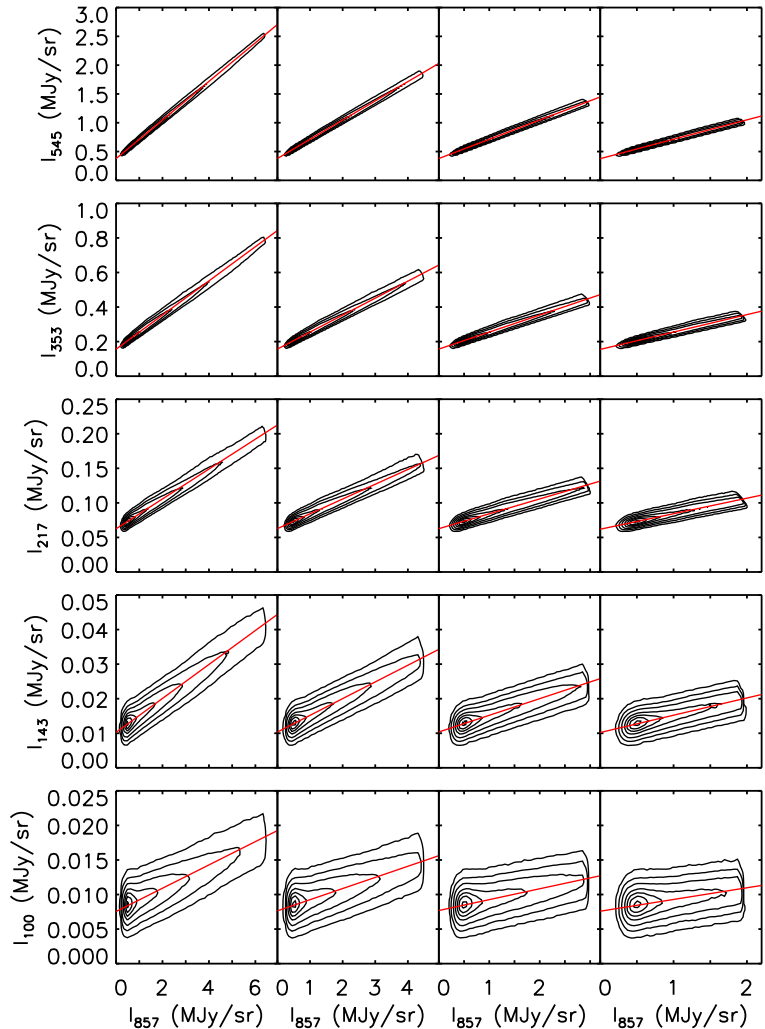}
\caption{Correlations of recalibrated HFI maps against the 857 GHz dust emission template.  Each row corresponds to a single 
frequency, and each column represents one of the four sky mask data selections (see text), with the largest sky area on the
left and the smallest area on the right.  The SMICA estimate of the CMB has been removed from the 100-353 GHz
data before fitting.  Contours show the density of points for individual pixels, and the contour levels are 1\%, 5\%, 15\%, 
35\%, 55\%, 75\%, and 95\% of the peak density for each panel.
The red line is the linear fit to the data.  Data are plotted on a scale designed to illustrate the stability of the fit
intercept with sky cut.
\label{fig:samplefit}
}
\end{figure*}

\begin{deluxetable*}{cccccc}
\tablecaption{Results of HFI vs. 857 GHz Dust Template Fits\tablenotemark{a} \label{tab:hfi_corr}}
\tablewidth{5in}
\tablehead{
\colhead{HFI Band} & 
\colhead{Slope $a_3$}  &
\colhead{$\sigma_{a_3}^{\rm{fit}}$ } &
\colhead{Intercept $c_3$} &
\colhead{$\sigma_{c_3}^{\rm{fit}}$ } &
\colhead{$\sigma_{c_3}^{\rm{sim}}$ }\\
\colhead{(GHz)} &
\colhead{} &
\colhead{} &
\colhead{(MJy sr$^{-1}$)} &
\colhead{(MJy sr$^{-1}$)} &
\colhead{(MJy sr$^{-1}$)}
}
\startdata
100  & 0.00161     &  0.00013     & 0.00766    &  0.00015     &  0.00016     \\
143  & 0.00480     &  0.00017     & 0.01042    &  0.00016     &  0.00032   \\
217  & 0.0214      &  0.0011      & 0.0630     &  0.0010      &  0.0010   \\
353  & 0.0972      &  0.0025      & 0.1586     &  0.0022      &  0.0015  \\
545  & 0.3344      &  0.0037      & 0.3923     &  0.0032      &  0.0036
\enddata
\tablenotetext{a}{Slopes and intercepts are the mean value over four different sky masks and three different CMB 
subtraction templates. $\sigma^{\rm{fit}}$ is the formal fitting uncertainty.  $\sigma_{c_3}^{\rm{sim}}$ is the
error in recovering the CIB monopole attributed to Galaxy removal systematics based on simulations.}
\end{deluxetable*}

\subsection{Correlations with HFI 857 GHz}

An alternative correlative analysis method utilizes a modified HFI 857 GHz map as the Galactic dust emission template.
The 857 GHz CIB value from the H~I and H$\alpha$ analysis is subtracted from the recalibrated, zodiacal light
subtracted 857 GHz map to form the template.
This serves as a more complete tracer of the ISM than H~I or H~I and H$\alpha$ templates.

For each of the 100 to 545 GHz HFI band maps, we perform a linear least-squares fit of the form 
\begin{equation}
I_\mathrm{HFI} = a_3 I_{857} + c_3
\end{equation}
where $a_3$ and $c_3$ are the outlier resistant slope and intercept, $I_{857}$ is the 857 GHz template,
and $I_\mathrm{HFI}$ represents recalibrated HFI map data from which zodiacal emission has been subtracted.
CMB anisotropies are also removed from the 100, 143, 217, and 353~GHz data, using three different
map estimates (SMICA, Commander, NILC) as a means of characterizing CMB removal error.
The CMB contributions to 545 and 857 GHz are ignored as subdominant.  All maps have been
smoothed to a common resolution of 16.2$\arcmin$ as described in \S3.1.

For the fitting, we adopt uncertainties appropriate for each frequency, including HFI measurement uncertainty 
calculated as described in \S2.1, CIB anisotropy noise estimated from simulations as described in \S3.1.2,
and uncertainty in the CMB anisotropy subtraction due to HFI measurement uncertainty, estimated from simulations. 
We make separate fits for four nested sky regions to assess stability of the results as a function of analyzed sky fraction. 
Each region is defined by the union of a diffuse emission 857 GHz intensity threshold, a point source mask for 857 GHz,
and a point source mask for the HFI band being fit.  The adopted point source masks are the same as those described in \S3.1.
Intensity cuts of  2.5, 3.5, 5, and 7 MJy/sr were selected for the recalibrated 857 GHz map prior to CIB subtraction.
The analyzed sky fraction is about 41\%, 52\%, 61\%, and 68\% for the four regions, although precise values vary with
the point source mask adopted for each
frequency.   A map of the four regions used for 545 GHz is shown in Figure 9(c). 

We fit for the slope $a_3$ and intercept $c_3$ for each sky cut and, where applicable, each case of CMB anisotropy subtraction. 
Figure 16 
illustrates the correlations and fitting results for all frequencies and sky cuts for the case 
of SMICA CMB removal. For each frequency, the results for the different sky cuts and CMB subtractions are in good agreement,
so we adopt the mean of the $a_3$ and $c_3$ values from the different fits. 
The rms scatter about the mean is adopted as the estimated fit uncertainty, $\sigma_{\rm{fit}}$. 
This is only approximate since results for different nested regions are not fully independent. In the case of the 
intercept, the quoted fit uncertainty also includes a contribution from the fit uncertainty for the adopted 857 GHz CIB 
from Table 7. The fit results are shown in Table 8. 

We use simulations to characterize potential bias in the fit intercept $c_3$ as a measure of the CIB due to any
large-scale imperfections in 
foreground removal resulting from spatial variations of the dust emission spectrum, or due to neglect of synchrotron and 
free-free foregrounds.  The latter effect is of little concern for HFI bands above 217 GHz. 

Each simulated sky realization contains Galactic emission, Gaussian noise, and CIB contributions, 
but assumes perfect 
CMB and zodiacal light removal.  The random number seed for the noise and CIB anisotropy generation is changed for each 
sky realization, while the Galaxy model remains unchanged. To simulate the Galactic emission in each HFI band, we 
use the {\it Planck} GNILC (Planck Collaboration Intermediate Results XLVIII 2016)
dust maps at 857, 545, and 353 GHz to model dust emission at these frequencies, but normalize their zero points to those of the
Meisner \& Finkbeiner (2015) Galactic thermal dust emission model.  For 100, 143, and 217 GHz, we use the Meisner \& Finkbeiner 
(2015) dust model itself, evaluated at the appropriate effective frequency for each band.  We evaluate the 2015 {\it Planck} Commander 
temperature component model (Planck Collaboration 2015 Results X 2016) to form maps of the contributions from synchroton and free-free
emission and add them to the dust maps at each frequency.

\begin{deluxetable*}{ccccccc}
\scriptsize
\tablewidth{0pt}
\tablecaption{Cosmic Infrared Background Results}
\tablehead{
\colhead{HFI Band} &
\colhead{CIB}&
\multicolumn{4}{c}{Contributions to CIB Uncertainty (MJy sr$^{-1}$)}&
\colhead{CIB Color}\\
\cline{3-6}
\colhead{(GHz)}&
\colhead{(MJy sr$^{-1}$)\tablenotemark{a}}&
\colhead{CMB Monopole}&
\colhead{Zodiacal Light}&
\colhead{HFI Offset}&
\colhead{Galactic Foreground}&
\colhead{Correction}\\
\colhead{}&
\colhead{}&
\colhead{Subtraction}&
\colhead{Subtraction}&
\colhead{Fit}&
\colhead{Subtraction}&
\colhead{}
}
\startdata
100 & $0.007 \pm 0.014$   & 0.011  & 0.00002 & 0.009 & 0.0005 & 1.076 \\
143 & $0.010 \pm 0.019$   & 0.017  & 0.00009 & 0.007 & 0.0006 & 1.017 \\
217 & $0.060 \pm 0.023$   & 0.022  & 0.0008  & 0.007 & 0.003  & 1.119 \\
353 & $0.149 \pm 0.017$   & 0.013  & 0.0045  & 0.008 & 0.006  & 1.097 \\
545 & $0.371 \pm 0.018$   & 0.0027 & 0.0095  & 0.012 & 0.009  & 1.068 \\
857 & $0.576 \pm 0.034$   & 0.0001 & 0.011   & 0.009 & 0.031  & 0.995
\enddata
\tablenotetext{a}{$\nu I_{\nu} (\mathrm{n\hskip -1pt W \: m}^{-2} \: \mathrm{sr}^{-1}) = \nu(\mathrm{GHz}) \, I_{\nu} (\mathrm{MJy \: sr}^{-1})/100$}
\end{deluxetable*}

\begin{figure*}
\figurenum{17}
\epsscale{0.9}
\vspace{0.5cm}
\plotone{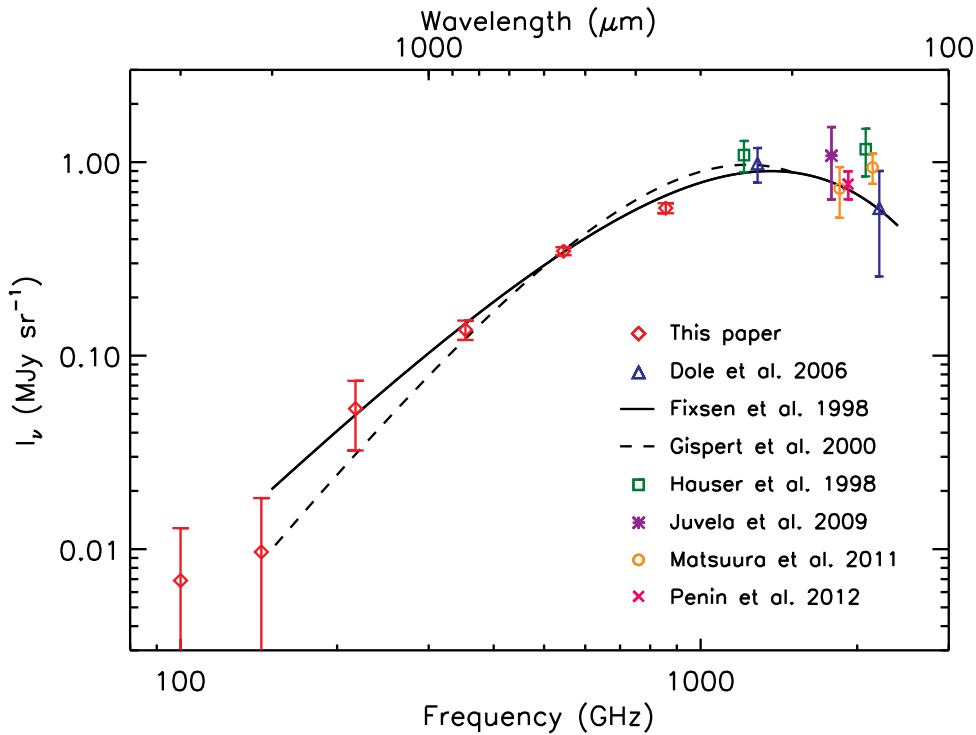}
\caption{Comparison of CIB results from this paper with previous direct CIB determinations at 
far-infrared to millimeter wavelengths. The diamonds show our color-corrected CIB results.
The solid curve shows the Fixsen et al. (1998) modified blackbody fit to
their CIB spectrum, and the dashed curve shows the modified blackbody fit
of Gispert et al. (2000) to the CIB spectrum determined by Lagache et al. (1999).
Other determinations shown are from analysis of {\it COBE}/DIRBE observations at 140 and
240 $\mu$m (Hauser et al. 1998), DIRBE observations transformed to the FIRAS photometric system 
(Dole et al. 2006), ISOPHOT observations at 150-180 $\mu$m (Juvela et al. 2009), AKARI observations
at 140 and 160 $\mu$m (Matsuura et al. 2011), and {\it Spitzer}/MIPS observations at 
160 $\mu$m (P{\'e}nin et al. 2012). Results from additional analyses of DIRBE data referred to in 
Section 1 are not shown but generally fall between the Hauser et al. and Dole et al. results.
Some of the data points are shown shifted slightly in frequency for clarity.}
\end{figure*}
 
Simulated maps for ten realizations are processed in the same way as the HFI data.
For each HFI band, the difference between the mean of the ten recovered fit intercepts and the known input CIB monopole is obtained
for each sky cut, and the mean of the absolute differences for the four cuts is taken as an estimate of 
the potential bias in 
$c_3$ as a measure of the CIB, $\sigma_{c_3}^{\rm{sim}}$.  This is listed in Table 8.
For each band, it is comparable to the fit uncertainty $\sigma_{c_3}^{\rm{fit}}$.

The intercepts from the 857 GHz template fits are consistent with our adopted CIB values from the H~I and H$\alpha$ fits
within the combined fit uncertainties and potential bias (Tables 7 and 8). The 857 GHz template method is not fully independent
since it uses the 857 GHz CIB from the H~I and H$\alpha$ fit, and any error in this value would be propagated to the
result for each of the 100 to 545 GHz bands. We prefer the H~I and H$\alpha$ fit results because they provide
independent CIB determinations for each HFI band.

\subsection{Summary of CIB Results}

Our adopted CIB values and uncertainties are listed in Table 9. Column 1 lists the CIB in each HFI band as
given by the intercept from our fit to the recalibrated HFI data using H~I and H$\alpha$. The total 
uncertainty listed in column 1 is the quadrature sum of the uncertainty contributions listed in columns 2
to 5. These are CMB monopole subtraction uncertainty from Table 1, zodiacal light subtraction uncertainty
from Table 3,
fit uncertainty for the HFI zero-level offset from Table 5, and Galactic foreground subtraction uncertainty as given
by the uncertainty in the fit intercept from Table 7. CMB subtraction uncertainty is the main contribution
for the 100 to 353 GHz bands. Galactic foreground subtraction uncertainty is the main contribution for the 857 
GHz band. For 545 GHz the main contributions are HFI offset uncertainty and Galactic foreground uncertainty.
The total CIB uncertainty is 11\%, 5\%, and 6\% of the CIB for the 353, 545, and 857 GHz bands,
respectively, significantly smaller than for previous direct CIB determinations in this frequency range
(Fixsen et al. 1998, Gispert et al. 2000). The CIB values are averages over the HFI bands as given
by equation (1). Column 6 gives color correction factors from Planck Collaboration 2013 Results XXX
(2014) for the CIB spectrum predicted by the B{\'e}thermin et al. (2012b) galaxy evolution model.

Our CIB results with these color corrections applied are compared to previous direct determinations
of the CIB at far infrared to millimeter wavelengths in Figure 17. Our results are consistent within
the uncertainties with previous determinations using FIRAS data by Fixsen et al. (1998) and 
Lagache et al. (1999). The uncertainties quoted for these previous results are $\sim30$\% or larger 
(see also Gispert et al. 2000), but we believe that they are underestimated at low frequencies because a 
contribution from CMB subtraction uncertainty was not included.

\section{Comparison with Source Counts}

\begin{deluxetable*}{cccccccl}
\footnotesize
\tablewidth{6.5in}
\tablecaption{Integrated Brightness of Selected Source Counts \label{tab:source_count_tab}}
\tablehead{
\colhead{} &
\colhead{} &
\colhead{Integrated} &
\colhead{Flux} &
\colhead{Extrapolated} &
\colhead{Survey} &
\colhead{Beam} &
\colhead{}\\
\colhead{$\lambda$} &
\colhead{$\nu$} &
\colhead{Brightness\tablenotemark{a}} &
\colhead{Limit} &
\colhead{Brightness} &
\colhead{Area} &
\colhead{FWHM} &
\colhead{Reference}\\
\colhead{($\mu$m)} &
\colhead{(GHz)} &
\colhead{(MJy sr$^{-1}$)} &
\colhead{($\mu$Jy)} &
\colhead{(MJy sr$^{-1}$)} &
\colhead{(arcmin$^2$)} &
\colhead{(arcsec)} &
\colhead{}.
}
\startdata
 1200 &  250 &$0.075^{+0.022}_{-0.018}$  &  20 &              &  9\tablenotemark{b}  &  $\lsim 1$  & ALMA; Fujimoto et al. (2016)  \cr
 1200 & 250 & $0.033 \pm 0.010$ & 200     &                      & 16\tablenotemark{b}  & $ \lsim 1$  & ALMA; Oteo et al. (2016) \cr
  850  & 353 & $0.068 \pm 0.009$\tablenotemark{c}&   100   &               &   ---    & $\sim 15$ &   SCUBA; Zemcov et al. (2010) \cr  
  850  & 353 & $0.122^{+0.069}_{-0.042}$ & 1000 &        &   646\tablenotemark{b}   & $\sim 14$ & SCUBA-2; Chen et al. (2013) \cr
  850  & 353 & $0.121^{+0.016}_{-0.013}$ &  100 &        &   1877\tablenotemark{b}   & $\sim 14$ & SCUBA-2; Hsu et al. (2016) \cr
  500  & 600 &                             &            &  $0.379 \pm 0.033$    & $\sim{2880}$\tablenotemark{d}    &    $60$           & BLAST; Marsden et al. (2009) \cr 
  500  & 600 &  $0.257 \pm 0.057$ &  2000  &  $0.47^{+0.16}_{-0.14}$     &  $\sim7200$\tablenotemark{e}  &  $36.6$      & {\it Herschel}/SPIRE; B{\'e}thermin et al. (2012a) \cr
  450  & 666 & $0.374^{+0.163}_{-0.093}$ & 1000 &        &  638\tablenotemark{b} & $7.5$  & SCUBA-2; Chen et al. (2013) \cr
  450  & 666 & $0.253^{+0.049}_{-0.044}$ & 1000 &        &  874\tablenotemark{b} & $7.5$  & SCUBA-2; Hsu et al. (2016) \cr
  450  & 666 & $0.295 \pm 0.056$  & 1000    &               & 151\tablenotemark{f}  & $\sim 8$ & SCUBA-2; Wang et al. (2017)  \cr
  350  & 857 & $0.53 \pm 0.11$ &  2000 &   $0.75^{+0.20}_{-0.18}$    &  $\sim7200$\tablenotemark{e}  & $24.9$    & {\it Herschel}/SPIRE; B{\'e}thermin et al. (2012a) \cr
  350  & 857 &                              &              &  $0.576 \pm 0.040$  & $\sim{2880}$\tablenotemark{d}   & $42$  & BLAST; Marsden et al. (2009) \cr
  250  & 1200 &                           &               &   $0.717 \pm 0.049$  & $\sim{2880}$\tablenotemark{d}   &  $36$ &   BLAST; Marsden et al. (2009) \cr
  250  & 1200 & $0.62 \pm  0.12$ &  2000 &   $0.84^{+0.22}_{-0.19}$    &  $\sim7200$\tablenotemark{e}  & $18.1$    & {\it Herschel}/SPIRE; B{\'e}thermin et al. (2012a) \cr
\enddata
\tablenotetext{a}{Includes both resolved and stacking analysis results, but not extrapolations. 1 MJy sr$^{-1}$ = 305 Jy deg$^{-2}$}
\tablenotetext{b}{Effective area over multiple pointings}
\tablenotetext{c}{Based on modeling of lensed source counts for 28 galaxy cluster fields. Error includes cosmic variance estimate}
\tablenotetext{d}{BLAST BGS-Deep field, which encompasses GOODS-S, Hubble Ultra Deep Field South, and Extended Chandra Deep Field South}
\tablenotetext{e}{COSMOS and GOODS-N fields}
\tablenotetext{f}{The smallest contiguous region included in the table. We estimate the cosmic variance of the mean CIB 
at 666 GHz for a region of this size to be 0.038 MJy sr$^{-1}$ (8.8\%), based on simulation of CIB anisotropy maps as described in section 3.1.2 but for a beam solid angle of 151 arcmin$^{2}$.}
\end{deluxetable*}

\begin{figure*}
\figurenum{18}
\begin{center}
\includegraphics[width=5.5in]{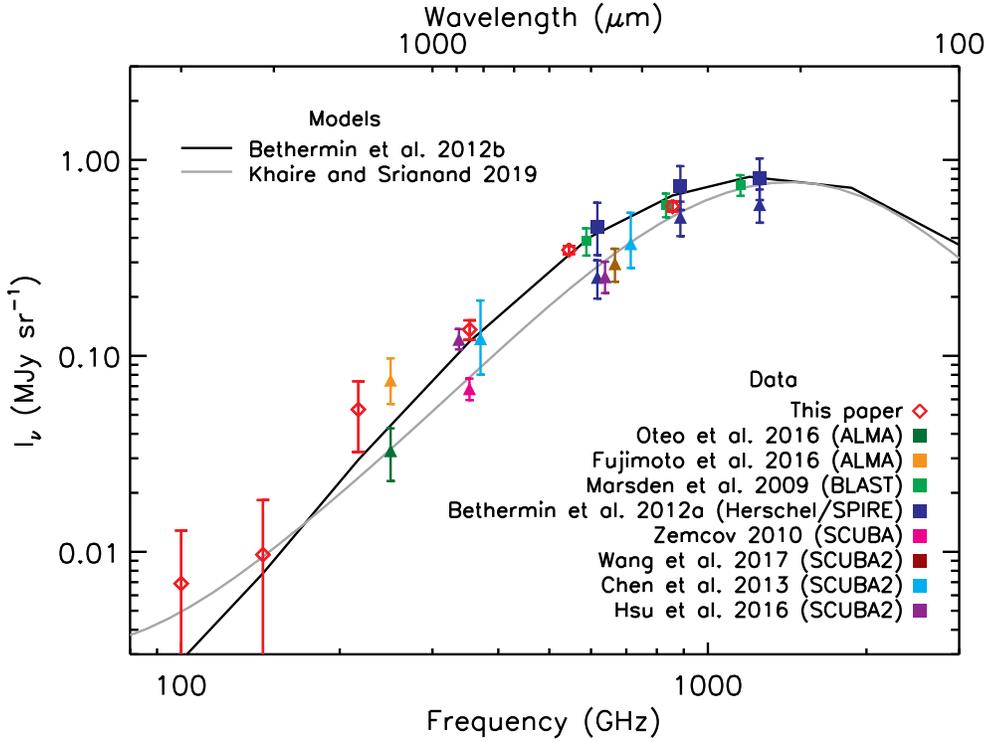}
\end{center} 
\vspace{-0.25cm}
\caption{Comparison of CIB results from this paper with published source count integrated brightness results. The
color-corrected results from this paper are shown as red open diamonds.  Selected source count results are color-coded as
labeled in the plot, with lower limits indicated by filled triangles, while extrapolations are shown as filled squares.  
Some of the source count results are shown shifted slightly in frequency for clarity. The black curve shows
predictions from the model of B{\'e}thermin et al. (2012b) in the {\it Planck} HFI, {\it Herschel}/SPIRE, and 
{\it Herschel}/PACS bands with color corrections from Planck Collaboration 2013 Results XXX (2014) and Maniyar et al. (2018) 
applied. The gray curve is from the physical modeling of Khaire and Srianand (2019).
\label{fig:src_cnt_compare}
}
\end{figure*}

Source number counts provide a key input toward
understanding and modeling galaxy formation and evolution.  A common 
metric is the cumulative contribution of integrated source intensities to the
CIB as a function of source flux and frequency. However,
the fraction of the CIB that may be attributed to resolved extragalactic sources is as yet uncertain.
At its most basic level, the computation requires accurate source counts over a wide flux limit range.  
Observational impediments in obtaining these source counts include source confusion,
instrument sensitivity limitations, and small sample sizes (cosmic variance).
With the advent of improved spatial resolution and sensitivity from observations such as those from 
ALMA and SCUBA-2, estimates of integrated source contributions to the CIB have been evolving rapidly 
in the literature.

We list a sample of published integrated source count results in Table~\ref{tab:source_count_tab}.
This is by no means a comprehensive list. 
In addition to selecting for those results at frequencies within the HFI spectral range,
we also have attempted to quote results in the literature that are expected to be least affected by
bias introduced from cosmic variance (sample variance).   For a contiguous sky region, 
this requires a fairly large observed sky area and, for example, excludes several of the newer
ALMA results based on analysis of a single area of a few square arcminutes or less.
We include recent results from ALMA that have adopted the approach of analyzing multiple 
spatially separated pointings in order to reduce bias from cosmic variance.  In these cases, as noted 
by e.g. Hsu et al. (2016), small scale clustering may still be a concern for bias.
In the table, we list results for resolved source counts integrated to a flux limit, including stacking methods.
These serve as lower limits to the CIB.
When provided, results are also listed from the authors' extrapolation of the number counts to lower flux 
based on expected source count turnover and a fitting function.

Number counts using newer sensitive high-resolution observations
have been shown to be in tension with older, lower-resolution surveys of the same sky region.
Wang et al. (2017) note that SCUBA-2 number counts
at 450 $\mu$m are shifted lower than those of {\it Herschel}/SPIRE at both 350 and 500 $\mu$m,
and hypothesize that source blending and/or clustering may have resulted in higher 
{\it Herschel}/SPIRE counts.  Simulations conducted by Bethermin et al. (2017) concur
with this conclusion, finding  that the number counts measured by {\it Herschel} at 350 and 500 $\mu$m 
between 5 and 50 mJy are biased toward high values roughly by a factor of 2. 

Discrepancies can also exist among interferometric determinations.
The ALMA number counts derived by Fujimoto et al. (2016) and Oteo et al. (2016) disagree by a factor of two at the same flux 
level.  Oteo et al. (2016) suggest differing detection limits and treatment of flux 
boosting (noise bias) as reasons for the discrepancy.  Hatsukade et al. (2018) compile cumulative source counts from their own ALMA survey
sample and several published ALMA results at 870 $\mu$m and 1.2~mm  and derive a Schechter function fit at 1.2~mm with associated
1$\sigma$ uncertainty.  Although they find that the Oteo et al. numbers to be lower than their mean values, Hatsukade et al. find the Oteo et al. number 
counts are overall consistent with their results within errors.  There is no indication of a faint-end turnover of these source counts out to their 
compilation limit  of 0.02 mJy, which is set by the Fujimoto et al. (2016) value in Table~\ref{tab:source_count_tab}.

In Figure 18, we show our direct CIB determinations along with the source count integrated brightness results compiled in 
Table~\ref{tab:source_count_tab}. 
For comparison purposes, we also show the model predictions of B{\'e}thermin et al. (2012b) and Khaire and Srianand (2019). Our CIB results are most
useful for comparison with integrated source count brightness results from 353 to 857 GHz, where the fractional CIB uncertainties are lowest.
We compare the SCUBA and SCUBA-2 results with our CIB results. We estimate the CIB at 666 GHz (450 $\mu$m) to be 0.44 MJy sr$^{-1}$ from a
modified blackbody fit to our CIB results and the 140 to 240 $\mu$m data points in Figure 17, 
$I_{\nu} = 2.73 \times 10^{-6} \thinspace (\nu /\mathrm{500 \thinspace GHz})^{0.315} \thinspace B_{\nu}(23.3 \thinspace \mathrm{K})$,
and we adopt an uncertainty of 6\%.  
For the ratio of integrated source brightness to CIB at 666 GHz, we obtain values of $85^{+37}_{-22}$\%, $58^{+12}_{-11}$\%, and 67\% $\pm$ 13\%
using the results of Chen et al. (2013), Hsu et al. (2016), and Wang et al. (2017), respectively.
For the ratio at 353 GHz, we obtain 50\% $\pm$ 9\%, $90^{+52}_{-33}$\%, and $89^{+15}_{-14}$\% using the results of Zemcov et al. (2010), 
Chen et al. (2013), and Hsu et al. (2016), respectively. Most of the quoted uncertainties are dominated by the uncertainty of the
integrated source brightness, and in all cases this contribution is larger than the contribution of the CIB uncertainty.
Accuracy of the resolved CIB fraction at these frequencies will improve with future more sensitive 
imaging surveys covering larger sky regions.

\section{Summary}

We have made new direct determinations of the cosmic infrared background monopole 
brightness in the {\it Planck} HFI frequency bands. In the 353 to 857 GHz bands,
our uncertainties are about 3 to 6 times smaller than those of previous CIB determinations
at these frequencies.

Correlation of HFI data from the 2015 release with FIRAS data was used to 
recalibrate the HFI map zero levels in all bands and to recalibrate the gains 
of the HFI 545 and 857 GHz band maps (Table 5). For the 100 to 353 GHz HFI 
bands the brightness scales of the HFI and FIRAS data were found to be consistent,
but the correlation with FIRAS data does not improve on the accuracy of the {\it Planck} team 
HFI gain calibration.
 
Correlation of the recalibrated HFI map for each band with different templates
of Galactic foreground emission was used for separating the CIB monopole brightness 
from the Galactic foreground. We obtained separate results using an H~I template, 
a combination of H~I and H$\alpha$ templates, and, for the 100 to 545 GHz bands,
a dust emission template formed from the recalibrated HFI 857 GHz map. CIB results
for the three cases were found to be consistent within the uncertainties. We adopted
the results from using H~I and H$\alpha$ (Tables 7 and 9) because an H$\alpha$ correlated 
component is detected that traces dust emission from the ionized medium and because 
this method gives an independent CIB determination for each HFI band. 
Due to a shortcoming of H$\alpha$ as a tracer of dust emission, the CIB brightness 
determined from correlation with H~I and H$\alpha$ may be somewhat overestimated, 
but we estimate the possible effect on our results to be smaller than 
our adopted CIB uncertainties.  

Our CIB uncertainties include contributions from uncertainty
in the HFI zero-level recalibration and uncertainties in subtraction of the CMB monopole, 
zodiacal light, and Galactic foreground emission and are as small as 5\% to 11\% for the 353
to 857 GHz bands.

The H$\alpha$ correlated emission has a spectral shape consistent with that of the H~I correlated 
emission.
The dust emissivity per H ion that we infer for the ionized medium is much lower
than the dust emissivity per H atom for the neutral atomic medium (Figure 15). Available evidence
suggests that this is at least partly due to a lower dust-to-gas mass ratio in the ionized medium
and partly due to the shortcoming of H$\alpha$ as a tracer of dust emission.
Error in the conversion of the H$\alpha$ correlation coefficient to dust emissivity per H ion may 
also be a factor.

Comparison of our direct CIB measurements with the integrated brightness of sources from recent
submillimeter to millimeter wavelength imaging surveys shows that a large fraction of the 
CIB, $\sim50$\% or more, has been resolved. The accuracy of the comparison in the 353 to 857 GHz
(350 to 850 $\mu$m) range
is limited by the accuracy of the source measurements but is expected to improve as
more sensitive imaging surveys covering larger sky regions are made.

\bigskip
 
We thank the referee and R. G. Arendt for helpful comments. We thank K. Ganga for help with 
questions about {\it Planck} team zodiacal light modeling. We thank 
L.-Y. Hsu for providing integrated source brightness values from the Hsu et al. (2016) SCUBA-2 
lensing cluster survey. This research was supported by the NASA Astrophysics Data Analysis 
Program, proposal 14-ADAP14-0114.

\appendix

\renewcommand{\theequation}{\arabic{equation}}

\section{CMB Monopole Subtraction Uncertainty}

We estimate the uncertainty in the best-fit CMB temperature we obtained for the pass 4 FIRAS data
in \S2.2.1 as the quadrature sum of contributions of measurement uncertainties, uncertainty in the dust
emission spectrum, and uncertainty in the CIB monopole spectrum subtraction that was applied before 
the fitting.

The contribution of measurement uncertainties was calculated by making fits to simulated datasets
that included the CMB monopole, Galactic dust emission, and different realizations of noise in the
FIRAS channels and DIRBE bands.
A separate simulation was done for each dust model, using the CMB temperature and dust model 
parameters from the fit to the observed data to obtain the simulation sky brightness spectrum 
for each sky region used in the fitting. Each noise realization was calculated by drawing
random deviates from a multivariate normal distribution with zero mean and covariance 
matrix described by Odegard et al. (2016), which includes DIRBE absolute
calibration gain and offset uncertainties and the following FIRAS uncertainties: detector noise, 
destriper uncertainty ($\beta$), bolometer model gain uncertainty (JCJ gain), calibration model
emissivity gain uncertainty (PEP gain), and internal calibrator temperature uncertainty (PUP).
For the MF dust model simulation, the standard deviation of recovered CMB temperature values for 
100 noise realizations is 5.7 $\mu$K. For the TLS dust model simulation, it is 8.2 $\mu$K for 100
realizations.  We adopt 
7 $\mu$K for the CMB temperature uncertainty due to measurement uncertainties. 

For the simulation
using the MF dust model, the average recovered CMB temperature is 6 $\mu$K greater
than the input CMB temperature. This bias is due to degeneracy between the CMB temperature
and other fit parameters for this model, especially with the emissivity index of the 
cold dust component, and we have applied a 6 $\mu$K bias correction to the
best-fit CMB temperature from our fit to the observed data using this model. After this correction, the 
difference between the CMB temperatures using the MF and TLS models is 9 $\mu$K, and we adopt this as the
CMB temperature uncertainty due to uncertainty in the dust emission spectrum.  No significant CMB 
temperature bias is found for the simulation using the TLS model.

The contribution of uncertainty in CIB monopole subtraction was estimated by comparing fit results 
for the baseline case where the Fixsen et al. (1998) modified blackbody fit to their CIB spectrum was
subtracted from the FIRAS data with results for the case where the Gispert et al. (2000) modified 
blackbody fit to the CIB spectrum of Lagache et al. (2000) was subtracted instead. As shown in Figure 17,
the CIB spectrum of Fixsen et al.
has greater amplitude and flatter slope than that of Gispert et al. at frequencies near the 160 GHz 
peak of the CMB spectrum. We found that using the Gispert et al. CIB spectrum results in a best-fit
CMB temperature that is 44 $\mu$K greater than that obtained using the Fixsen et al. CIB spectrum,
and we found essentially the same result whether the MF dust model or the TLS dust model was used
in the fitting.  We adopt 44 $\mu$K as the uncertainty in the best-fit CMB temperature due to 
uncertainty in the CIB monopole subtraction. Adding it in quadrature with the contributions of measurement 
uncertainties and uncertainties in the dust spectrum, we obtain a combined uncertainty for the best 
fit CMB temperature of 46 $\mu$K.

We have found that contributions of uncertainty in subtraction of zodiacal light, free-free emission,
and synchrotron emission before the fitting are negligible. 
For our baseline fitting, free-free and synchrotron maps from the WMAP 9-year maximum entropy method
foreground separation were extrapolated in frequency and subtracted from the FIRAS data.  We found that the 
best-fit CMB temperature changes by less than 1 $\mu$K if instead no free-free subtraction or synchrotron 
subtraction is applied. 
For zodiacal light subtraction, our baseline fitting used (1) the {\it Planck} team 2015 zodiacal light model evaluated
as described in \S2.2.4 at the HFI band frequencies and then interpolated to the FIRAS channel frequencies for 
$100 < \nu \leq 857$ GHz, (2) the FIRAS zodiacal light model for $\nu \geq 1250$ GHz, and (3) interpolation
between these model predictions for the channels between 857 and 1250 GHz. We found that the best-fit CMB 
temperature changes by less than 1 $\mu$K if instead the FIRAS zodiacal light model is used for all frequencies.

\section{FIRAS Noise Covariance Matrix} \label{app:covar}

\setcounter{equation}{9}

We follow the method described in section 7 of the FIRAS explanatory supplement (Brodd et al. 1997)
to calculate the noise covariance matrix for FIRAS data averaged over the FIRAS frequency channels in 
a given HFI band. This is a 6063 x 6063 matrix, where 6063 is the number of FIRAS pixels that were 
observed in both the low frequency and high frequency bands over the {\it COBE} mission. We calculate the covariance matrix as a sum of contributions
from the following FIRAS uncertainties: detector noise, destriper uncertainties ($\beta$, which 
include offset uncertainties), bolometer model gain uncertainties (JCJ gain), calibration model 
emissivity gain uncertainties (PEP gain), and internal calibrator temperature uncertainty (PUP).  Uncertainty in the
absolute temperature scale of the FIRAS external calibrator (PTP) is not included, since it is only
important in determination of the absolute temperature of the CMB and we have treated this as a
nuisance parameter.

The channel-averaged FIRAS intensity for a given pixel is expressed as the result of a
linear operator $H_{pi}$ acting on the FIRAS data,
\begin{equation}
I^{FIRAS}_{\nu_0, p} = H_{pi}(F^{pi}) = \frac{\sum\limits_{i} \thinspace F^{pi} A_i}{\sum\limits_{i} \thinspace (\nu_0/\nu_i) A_i}.
\end{equation}
Here $F^{pi}$ is the FIRAS destriped sky spectrum data after CMB monopole subtraction for pixel $p$ and 
frequency channel $i$, $\nu_0$ is the nominal 
HFI band frequency, $\nu_i$ is the center frequency for channel $i$, and
$A_i$ is the channel response function fit amplitude for channel $i$.
Following are expressions for terms in the FIRAS covariance matrix for this operator that follow
from equation (35) of the FIRAS explanatory supplement.

The contribution of detector noise to the covariance matrix is
\begin{equation}
H_{pi}H_{p'i'}(C^{ii'}\delta^{pp'}/N_p) = \frac{\delta^{pp'}}{N_p} \left [\frac {1}{\sum\limits_{i} \thinspace (\nu_0/\nu_i) A_i} \right ]^2 {\sum\limits_{i} A_i} {\sum\limits_{i'} A_{i'}C^{ii'}},
\end{equation}
where $C^{ii'}$ is the FIRAS C matrix described in section 7.1.2 of the FIRAS explanatory supplement, 
$\delta^{pp'}$ is the Kronecker delta function indicating that off-diagonal elements are zero, and $N_p$ is the FIRAS pixel weight.

The contribution of destriper uncertainties is
\begin{equation}
H_{pi}H_{p'i'} (\beta^p_k \beta_{p'k} + 0.04^2) C^{ii'} = (\beta^p_k \beta_{p'k} + 0.04^2) \left [\frac {1}{\sum\limits_{i} \thinspace (\nu_0/\nu_i) A_i} \right ]^2 {\sum\limits_{i} A_i} {\sum\limits_{i'} A_{i'}C^{ii'}},
\end{equation}
where $\beta^p_k$ is the beta matrix described in section 7.2.2 of the FIRAS explanatory supplement 
and we sum $\beta^p_k \beta_{p'k}$ over all orthogonalized stripes $k$.

The contribution of JCJ gain uncertainties is
\begin{equation}
H_{pi}H_{p'i'}(S^{pi}S^{p'i'}J^iJ^{i'}) = \left [\frac {1}{\sum\limits_{i} \thinspace (\nu_0/\nu_i) A_i} \right ]^2 {\sum\limits_{i} A_i S^{pi} J^{i}} {\sum\limits_{i'} A_{i'} S^{p'i'} J^{i'}}
\end{equation}
where $S^{pi}$ is the absolute sky brightness not including the CMB monopole and $J^i$ is the 
JCJ gain term described in section 7.3.2 of the FIRAS explanatory supplement. The FIRAS data $F^{pi}$ 
could be used for $S^{pi}$ but they have low signal-to-noise ratio in the lower frequency channels.
We have chosen to use instead the smoothed HFI data with zodiacal light 
and CMB dipole contributions added, scaled to each FIRAS channel in a given HFI band using the spectral model
described in section 2.2.5 for the CIB monopole, CMB fluctuation, and Galactic emission components, together with the
model zodiacal light spectrum and the CMB dipole spectrum.

The contribution of PEP gain uncertainties is
\begin{equation}
H_{pi}H_{p'i'}(S^{pi}S^{p'i'}G^iG^{i'}\delta^{ii'}) = \left [\frac {1}{\sum\limits_{i} \thinspace (\nu_0/\nu_i) A_i} \right ]^2 {\sum\limits_{i} A_{i}^2 (S^{pi} S^{p'i} G^{i} G^{i})}
\end{equation}
where $G^i$ is the PEP gain term described in section 7.2.3 of the FIRAS explanatory supplement.

The contribution of the PUP uncertainty is
\begin{equation}
H_{pi}H_{p'i'}(P^{i}P^{i'}U^2\delta^{pp'}/N_p) = \frac{U^2 \delta^{pp'}}{N_p} \left [\frac {\sum\limits_{i} A_i P^{i}}{\sum\limits_{i} \thinspace (\nu_0/\nu_i) A_i}\right ]^2
\end{equation}
where $P^{i}$ is $\partial B(2.728 K,\nu_i)/\partial T$ and $U$ is the internal calibrator temperature uncertainty. We 
take $U$ to be 150 $\mu$K, the recommended estimate from section 7.4.5 of the FIRAS explanatory supplement.

\section{H$\alpha$ as a Tracer of Dust Emission}

\setcounter{equation}{15}

The issue with using H$\alpha$ as a tracer of dust emission from the ionized medium is that the
intensity of this emission varies in proportion to dust column density, but H$\alpha$ intensity
varies in proportion to the square of the ionized gas density integrated along the line of sight, 
$I$(H$\alpha$) $\propto T_e^{-0.92} \int n_e^2 ds$, where $n_e$ is electron density
and $T_e$ is electron temperature (Reynolds 1992). Thus, errors would be expected in our results
if the variation of H$\alpha$ intensity in the regions we analyze is caused more by 
variation in mean electron density or mean electron temperature than by variation in ionized 
gas column density.
 
Odegard et al. (2007) considered a simple model in which the fractional H$\alpha$ variation in a 
sky region is represented as the product of the fractional variation of $N$(H$^{+}$) and the
fractional variation of effective electron density, $n_{eff} \equiv \int n_e^2 ds/\int n_e ds$,
assuming that electron temperature is constant.
They adopted $N$(H$^{+}$) $\propto I$(H$\alpha$)$^{p}$ and $n_{eff} \propto I$(H$\alpha$)$^{(1-p)}$,
where $p$ is the fraction of the variation of log $I$(H$\alpha$) that is caused by variation of $N$(H$^+$).
If the distributions of $N$(H~I) and $I$(H$\alpha$) are uncorrelated, they showed for this model that
a correlation analysis using H~I and H$\alpha$ would underestimate the emissivity of the ionized medium,
\begin{equation}
\epsilon_{derived} (\rm H^{+}) = p \thinspace \epsilon (\rm H^{+}) ,
\end{equation}
and would overestimate the CIB by
\begin{equation}
\Delta I_{\nu} = \frac {(1-p)} {p} \thinspace b \thinspace \langle I(\rm H\alpha) \rangle .
\end{equation}
Here $b$ is the H$\alpha$ fit coefficient (equation 8) and $\langle I(\rm H\alpha) \rangle$ is the
mean H$\alpha$ brightness for the sky region. 
We use this to estimate the amount by which our H~I and H$\alpha$ fit intercept may exceed the CIB in each of the HFI bands.  

We adopt $p = 0.67$ based on studies of the correlation between dispersion measure DM and emission 
measure for lines of sight to pulsars. Berkhuijsen and M\"uller (2008) obtained $p = 0.87 \pm 0.05$
using dispersion measure data for a sample of 34 pulsars at $\vert b \vert \gtrsim 4\arcdeg$
with distances known better than 50\%. They used emission measures toward these pulsars from the 
H$\alpha$ map of Finkbeiner (2003), assuming an electron temperature of 8000 K and applying 
corrections for extinction and for the contribution to emission measure originating from beyond 
each pulsar. 
Pynzar' (2016) obtained $p = 0.500 \pm 0.005$ using a sample of 120 pulsars chosen to have 
$18 < {DM} \: \mathrm{sin} \vert b \vert < 27$ pc cm$^{-3}$, to select pulsars likely to be at high $z$
and to exclude pulsars in directions of H~II regions.  Emission measures were obtained
from background brightness temperature measurements at 1420, 2695, 5000, and 10550 MHz 
corrected for the contribution of nonthermal emission, H166$\alpha$ recombination
line intensity, H$\alpha$ intensity, or estimated thermal background brightness temperature at
408 MHz. The Berkhuijsen and M\"uller result may be more relevant here since the latitude
distribution of their pulsar sample is in better agreement with our sky regions. The Pynzar'
sample includes many lines of sight in the Galactic plane.
Our adopted value of $p$ bisects the correlation slopes from the two studies.

Using this value of $p$, values of $b$ from our fits for the $N_\mathrm{HI}$ cut = $2.5 \times 10^{20}$ 
cm$^{-2}$ region from Table 7, and the mean extinction-corrected and scattering-corrected H$\alpha$ 
intensity of 0.86 R for this region, equation (17) gives $\Delta I_{\nu}$ of
0.00006, 0.00015, 0.00068, 0.0032, 0.011, and 0.028 MJy sr$^{-1}$ at 100, 143, 217, 353, 545, and 857 GHz, 
respectively. Given the model assumptions and the uncertainties of the parameter $p$, these are only rough 
estimates of the amount by which our fit intercepts may exceed the CIB, and we choose to adopt 
the fit intercepts as CIB values without applying any corrections. At 857 GHz, the estimated
$\Delta I_{\nu}$ is 0.8 times the total CIB uncertainty listed in Table 9, and for the 
other frequencies $\Delta I_{\nu}$ as a fraction of the CIB uncertainty is smaller.


\newpage

\begin{references}

\reference{} Arendt, R. G., Odegard, N., Weiland, J. L., et al. 1998, ApJS, 508, 74

\reference{} Bennett, C.L., Larson, D., Weiland, J. L., et al., 2013, ApJS, 208, 20B

\reference{} Berkhuijsen, E. M., \& M\"uller, P. 2008, \aap, 490, 179

\reference{} B{\'e}thermin, M., Le Floc'h, E., Ilbert, O., et al. 2012a, \aap, 542, A58

\reference{} B{\'e}thermin, M., Daddi, E., Magdis, G., et al. 2012b, ApJ, 757, L23

\reference{} B{\'e}thermin, M., Wu, H.-Y., Lagache, G., et al. 2017, \aap, 607, 89

\reference{} Boulanger, F., \& Perault, M. 1988, \apj, 330, 964

\reference{} Boulanger, F., Abergel, A., Bernard, J.-P., et al. 1996, \aap, 312, 256

\reference{} Brandt, T. D., \& Draine, B. T. 2012, ApJ, 744, 129

\reference{} Brodd, S., Fixsen, D. J., Jensen, K. A., Mather, J. C., \&
Shafer, R. A. 1997, {\it COBE} Far Infrared Absolute Spectrophotometer (FIRAS)
Explanatory Supplement, {\it COBE} Ref. Pub. No. 97-C (Greenbelt, MD: NASA/GSFC),
available in electronic form from https://lambda.gsfc.nasa.gov

\reference{} Casey, C. M., Narayanan, D., \& Cooray, A. 2014, Physics Reports, 541, 45

\reference{} Chen, C.-C., Cowie, L. L., Barger, A. J., et al. 2013, ApJ, 776, 131

\reference{} Cooray, A. 2016, Royal Society Open Science, 3, 150555

\reference{} Deul, E. R., \& Burton, W. B. 1990, \aap, 230, 153

\reference{} Deul, E. R., \& Burton, W. B. 1992, in The Galactic Interstellar
Medium, ed. D. Pfenniger and P. Bartholdi, (Heidelberg: Springer-Verlag), p. 79

\reference{} Dole, H., Lagache, G., Puget, J.-L., et al. 2006, A \& A, 451, 417

\reference{} Dwek, E. \& Krennrich, F. 2013, Astroparticle Physics, 43, 112

\reference{} Finkbeiner, D. P. 2003, ApJS, 146, 407

\reference{} Finkbeiner, D. P., Davis, M., \&  Schlegel, D. J., 1999, \apj, 524, 867

\reference{} Finkbeiner, D. P., Davis, M., \&  Schlegel, D. J., 2000, \apj, 544, 81

\reference{} Fixsen, D. J., Cheng, E. S., Gales, J. M., et al. 1996, ApJ, 473, 576

\reference{} Fixsen, D. J., Weiland, J. L., Brodd, S., et al. 1997, \apj, 490, 482

\reference{} Fixsen, D. J., Dwek, E., Mather, J. C., Bennett, C. L.,
\& Shafer, R. A. 1998, \apj, 508, 123

\reference{} Fixsen, D. J. 2009, \apj, 707, 916

\reference{} Fujimoto, S., Ouchi, M., Ono, Y., et al. 2016, ApJS, 222, 1 

\reference{} Gaensler, B. M., Madsen, G. J., Chatterjee, S., \& Mao, S.A. 2008, PASA, 25, 184

\reference{} Gispert, R., Lagache, G., \& Puget, J. L. 2000, \aap ,360, 1

\reference{} Gorjian, V., Wright, E.L., \& Chary, R.R. 2000, \apj, 536, 550

\reference{} G{\'o}rski, K. M., Hivon, E., Banday, A. J., et al. 2005, ApJ, 622, 759

\reference{} Haffner, L. M. 2001, in Tetons 4: Galactic Structure, Stars, and the Interstellar Medium,
  ASP Conference Series, Vol. 231, ed. C. E. Woodward, M. D. Bicay, and J. M. Shull, 345
  (San Francisco: Astronomical Society of the Pacific)

\reference{} Haffner, L. M., Reynolds, R. J., \& Tufte, S. L. 1999, \apj, 523, 223

\reference{} Haffner, L. M., Reynolds, R. J., Tufte, S. L., et al. 2003, ApJ, 149, 405

\reference{} Haffner, L. M., Reynolds, R. J., Madsen, G. J., et al. 2010, in
  The Dynamic Interstellar Medium: A Celebration of the Canadian Galactic Plane Survey,
  ASP Conference Series, Vol. 438, ed. R. Kothes, T. L. Landecker, and A. G. Willis, 388
  (San Francisco: Astronomical Society of the Pacific)

\reference{} Haffner, L. M., Reynolds, R. J., Madsen, G. J., et al. 2018, in preparation

\reference{} Hatsukade, B., Kohno, K., Yamaguchi, Y., et al. 2018, PASJ, 70, 105

\reference{} Hauser, M. G., Arendt, R. G., Kelsall, T., et al. 1998, \apj, 508, 25

\reference{} Hauser, M. G., \& Dwek, E. 2001, ARA\&A, 39, 249

\reference{} HI4PI Collaboration 2016, \aap, 594, A116

\reference{} Hinshaw, G., Weiland, J. L., Hill, R. S., et al. 2009, ApJS, 180, 225

\reference{} Howk, J. C., \& Savage, B. D. 1999, \apj, 517, 746

\reference{} Howk, J. C., Sembach, K. R., \& Savage, B. D. 2003, \apj, 586, 249

\reference{} Hsu, L.-Y., Cowie, L. L., Chen, C.-C., et al. 2016, ApJ, 829, 25

\reference{} Jenkins, E. B. 1987, in Interstellar Processes, ed. D. J. Hollenbach and
H. A. Thronson, Jr. (Dordrecht: Reidel), 533

\reference{} Juvela, M., Mattila, K., Lemke, D., et al. 2009, \aap, 500, 763

\reference{} Kalberla, P. M. W., and Haud, U. 2015, \aap, 578, 78

\reference{} Kashlinsky, A. 2005, Physics Reports, 409, 361

\reference{} Kelsall, T., Weiland, J. L., Franz, B. A., et al. 1998, \apj, 508, 44

\reference{} Khaire, V., and Srianand, R. 2019, MNRAS, 484, 4174

\reference{} Kondo, T., Ishihara, D., Kaneda, H., et al. 2016, AJ, 151, 71

\reference{} Lagache, G., Abergel, A., Boulanger, F., et al. 1999, \aap, 344, 322

\reference{} Lagache, G., Haffner, L. M., Reynolds, R. J., \& Tufte, S. L. 2000, \aap, 354, 247

\reference{} Lagache, G., Puget, J.-L., \& Dole, H. 2005, ARA\&A, 43, 727

\reference{} Mak, D. S. Y., Challinor, A., Efstathiou, G., and Lagache, G. 2017, MNRAS, 466, 286

\reference{} Maniyar, A., B{\'e}thermin, M., and Lagache, G. 2018, \aap, 614, A39
 
\reference{} Marsden, G., Ade, P. A. R., Bock, J. J., et al. 2009, ApJ, 707, 1729

\reference{} Mather, J. C., Fixsen, D. J., Shafer, R. A., Mosier, C., 
\& Wilkinson, D. T. 1999, \apj, 512, 511

\reference{} Matsuura, S., Shirahata, M., Kawada, M., et al. 2011, \apj, 737, 2

\reference{} Meisner, A. M., and Finkbeiner, D. P. 2015, \apj, 798, 88

\reference{} Meny, C., Gromov, V., Boudet, N., et al. 2007, \aap, 468, 171

\reference{} Odegard, N., Arendt, R. G., Dwek, E., Haffner, L. M., Hauser, M. G.,
\& Reynolds, R. J. 2007, \apj, 667, 11

\reference{} Odegard, N., Kogut, A., Chuss, D. T., \& Miller, N. J. 2016, \apj, 828, 16

\reference{} Oteo, I., Zwaan, M. A., Ivison, R. J., et al. 2016, \apj, 822, 36

\reference{} Paradis, D., Bernard, J.-Ph., Meny, C., \& Gromov, V. 2011, \aap, 534, 118

\reference{} P{\'e}nin, A., Lagache, G., Noriega-Crespo, A., et al. 2012, \aap, 543, A123

\reference{} Planck Collaboration Early Results XVIII 2011, \aap, 536, A18

\reference{} Planck Collaboration Early Results XXIV 2011, \aap, 536, A24

\reference{} Planck Collaboration 2013 Results VIII 2014, \aap, 571, A8

\reference{} Planck Collaboration 2013 Results XI 2014, \aap, 571, A11

\reference{} Planck Collaboration 2013 Results XIV 2014, \aap, 571, A14

\reference{} Planck Collaboration 2013 Results XXX 2014, \aap, 571, A30

\reference{} Planck Collaboration 2015 Results I 2016, \aap, 594, A1

\reference{} Planck Collaboration 2015 Results VIII 2016, \aap, 594, A8

\reference{} Planck Collaboration 2015 Results X 2016, \aap, 594, A10

\reference{} Planck Collaboration 2018 Results IV 2018, \aap, submitted, arXiv:1807.06208

\reference{} Planck Collaboration Intermediate Results XVII 2014, \aap, 566, A55

\reference{} Planck Collaboration Intermediate Results XLVI 2016, \aap, 596, A107

\reference{} Planck Collaboration Intermediate Results XLVIII 2016, \aap, 596, A109

\reference{} Press, W. H., Teukolsky, S. A., Vetterling, W. T., \&
Flannery, B. P. 1992, Numerical Recipes in FORTRAN, Second Edition,
(Cambridge: Cambridge University Press), 660 

\reference{} Puget, J.-L., Abergel, A., Bernard, J.-P., et al. 1996, \aap, 308, L5

\reference{} Pynzar', A. V. 2016, ARep, 60, 332

\reference{} Reach, W. T., Koo, B.-C., \& Heiles, C. 1994, \apj, 429, 672

\reference{} Reynolds, R. J. 1985, AJ, 90, 92

\reference{} Reynolds, R. J. 1991, in IAU Symp. 144, The Interstellar
Disk-Halo Connection in Galaxies, ed. H. Bloemen (Dordrecht: Kluwer), 67

\reference{} Reynolds, R. J. 1992, \apj, 392, L35

\reference{} Reynolds, R. J., Tufte, S. L., Kung, D. T., et al. 1995, \apj, 448, 715

\reference{} Reynolds, R. J., Chaudhary, V., Madsen, G. J., \& Haffner, L. M. 2005, 
 AJ, 129, 927

\reference{} Rowan-Robinson, M., \& May, B. 2013, MNRAS, 429, 2894

\reference{} Savage, B. D. \& Sembach, K. R. 1996, \araa, 34, 279

\reference{} Savage, B. D., and Wakker, B. P. 2009, \apj, 702, 1472

\reference{} Schlegel, D. J., Finkbeiner, D. P., \& Davis, M. 1998, \apj, 500, 525

\reference{} Wakker, B. P. 2004, in High Velocity Clouds, ed. H. van Woerden et al., 25

\reference{} Wang, W.-H, Lin, W.-C., Lim, C.-F., et al. 2017, ApJ, 850, 37

\reference{} Winkel, B., Kerp. J., Fl{\"o}er, L., et al. 2016, \aap, 585, A41

\reference{} Witt, A. N., Gold, B., Barnes, F. S., III, et al. 2010, ApJ, 724, 1551

\reference{} Wright, E. L. 2004, New Astr. Rev., 48, 465

\reference{} Zavala, J. A., Aretxaga, I., Geach, J. E., et al. 2017, MNRAS, 464, 3369

\reference{} Zemcov, M, Blain, A., Halpern, M., and Levenson, L. 2010, ApJ, 721, 424

\end{references}
\end{document}